\documentclass[twocolumn,showpacs,prl,superscriptaddress]{revtex4}
\usepackage{graphicx}
\usepackage{color}
\usepackage{amssymb,amsfonts,amsmath}
\usepackage{natbib}
\usepackage{subfigure}
\usepackage{epsfig}
\usepackage{epstopdf}
\usepackage{appendix}
\usepackage{comment}

\begin{document}

\title{Global topological control for synchronized dynamics on networks.}

\author{Giulia Cencetti} 
\affiliation{Dipartimento di Ingegneria dell'Informazione, Universit\`{a} di Firenze,
Via S. Marta 3, 50139 Florence, Italy}
\affiliation{INFN Sezione di Firenze, via G. Sansone 1, 50019 Sesto Fiorentino, Italia}
\author{Franco Bagnoli}  \affiliation{Universit\`{a} degli Studi di Firenze, Dipartimento di Fisica e Astronomia and CSDC, via G. Sansone 1, 50019 Sesto Fiorentino, Italia}
\affiliation{INFN Sezione di Firenze, via G. Sansone 1, 50019 Sesto Fiorentino, Italia}
\author{Giorgio Battistelli} 
\affiliation{Dipartimento di Ingegneria dell'Informazione, Universit\`{a} di Firenze,
Via S. Marta 3, 50139 Florence, Italy}
\author{Luigi Chisci} 
\affiliation{Dipartimento di Ingegneria dell'Informazione, Universit\`{a} di Firenze,
Via S. Marta 3, 50139 Florence, Italy}
\author{Francesca Di Patti} \affiliation{Universit\`{a} degli Studi di Firenze, Dipartimento di Fisica e Astronomia and CSDC, via G. Sansone 1, 50019 Sesto Fiorentino, Italia}
\affiliation{INFN Sezione di Firenze, via G. Sansone 1, 50019 Sesto Fiorentino, Italia}
\author{Duccio Fanelli} \affiliation{Universit\`{a} degli Studi di Firenze, Dipartimento di Fisica e Astronomia and CSDC, via G. Sansone 1, 50019 Sesto Fiorentino, Italia}
\affiliation{INFN Sezione di Firenze, via G. Sansone 1, 50019 Sesto Fiorentino, Italia}

\begin{abstract}
A general scheme is proposed and tested to control the symmetry breaking instability of a homogeneous solution of a spatially extended multispecies model, defined on a network. The inherent discreteness of the space makes it possible to act on the topology of the inter-nodes contacts to achieve the desired degree of stabilization, without altering the dynamical parameters of  the model. Both symmetric and asymmetric couplings are considered.  In this latter setting the  web of contacts is assumed to be balanced, for the homogeneous equilibrium to exist. The performance of the proposed method are assessed, assuming the Complex Ginzburg-Landau equation as a reference model. In this case, the implemented control allows one to stabilize the synchronous limit cycle, hence time-dependent, uniform solution. A  system of coupled real  Ginzburg-Landau equations is also investigated to obtain the topological stabilization of a homogeneous and constant fixed point. 
\end{abstract}


\maketitle
\section{Introduction}

Self-organized collective dynamics may emerge  in systems constituted by many-body interacting entities \cite{CrossHohenberg93, Hoyle06}. This is a widespread observation in nature which fertilized in a cross-disciplinary perspective to ideally embrace distinct  realms of investigations. Convection instabilities in fluid dynamics, weak turbulences and defects are among the examples that  testify on the inherent ability of physical systems to yield coherent dynamical behaviors \cite{DrazinReid04}.  Insect swarms and fish schools exemplify the degree of spontaneous coordination that can be reached in  ecological applications \cite{Murray03}, while rhythms production and the brain functions refer to archetypical illustrations drawn from biology and life science in general \cite{Goldbeter97, WyllerBlomquistEinevoll07, NakaoMikhailov10, AsllaniBiancalaniFanelliMcKane13, AsllaniDiPattiFanelli12, Strogatz01}. 
The mathematics  that underlies patterns formation focuses on the dynamical interplay between reaction and diffusion processes.  Usually, reaction-diffusion models are defined on a regular lattice, either continuous or discrete \cite{Murray02}. In many cases of interest, it is however more natural to schematize the system as a network, bearing a heterogeneous and complex structure \cite{OthmerScriven71, OthmerScriven74, NakaoMikhailov10, AsllaniChallengerPavoneSacconiFanelli14, AngstmannDonnellyHenry13}. To account for the hierarchical organization in multiple nested layers, networks of networks can be also considered \cite{AsllaniBusielloCarlettiFanelliPlanchon14, MuchaRichardsonMaconPorter10, GomezReinaresArenasFloria12, Bianconi13, MorrisBarthelemy12, NicosiaBianconiLatoraBarthelemy13, KivelaArenasBarthelemy14, Boccaletti_etal14, KouvarisHataDiaz15}. Imagine microscopic fluctuations  to shake a stationary stable, homogeneous equilibrium of the analyed, spatially extended, system. Under specific conditions, the imposed fluctuations get self-consistently enhanced by an intrinsic resonance mechanism,  which is ultimately triggered by the spatial component of the dynamics: instabilities seeded by random perturbations are often patterns precursors and eventually materialize in beautiful patchy motifs for the concentration of the mutually interacting species \cite{MimuraMurray78, MaronHarrison97, BaurmannGrossFeudel07, RietkerkVandeKoppel08, MeinhardtGierer00, HarrisWilliamsonFallon05, MainiBakerChuong06, NewmanBhat07, MiuraShiota00, WyllerBlomquistEinevoll07, ZhabotinskyDolnik95}. These are the byproduct of the celebrated Turing instability, named after Alan Turing who first conceived the symmetry breaking mechanism from which the process originates \cite{Turing52}. For reaction-diffusion systems hosted on a graph, the instability typifies a characteristic asymptotic segregation in rich (resp. poor) nodes of activators (resp. inhibitors),  a non homogeneous attractor that unavoidably emanates from the initially synchronous configuration \cite{NakaoMikhailov10, ContemoriDiPattiFanelliMiele16, PetitCarlettiAsllaniFanelli15}. 

In many cases of interest it is however important to oppose the natural drive to pattern formation, by preserving (or recovering) the synchronized state \cite{Pikovsky03, MirolloStrogatz90}.  Synchronization plays indeed a pivotal role in many branches of science:  the efficient coordination of a multitude of events is often decisive to have a system operated as a unison orchestra. In an alternating current electric power grid, one needs to match the speed and frequency of any given generator to the other sources of the shared network \cite{DorflerBullo12, DorflerChertkovBullo13, JungKettemann16}. In neuroscience,  patterns of synchronous firings are promoted by dedicated neuronal feedbacks. Circadian rhythms are another example that certifies the ubiquitous tendency towards entrainable oscillations as displayed by a vast plethora of biological processes \cite{Goldbeter97, Gandica_etal16}. In computer science, synchronization is customarily referred to as consensus \cite{OlfatiFaxMurray07}, a form of final agreement, stationary or time dependent, which is reached by a crowd of interacting agents.

Given these premises it is in general important to devise apt control strategies that enable one to stabilize, and possibly preserve, the synchronous regime. The control is classically applied to the reactive component of the dynamics, and ultimately shape the local interaction between constitutive elements \cite{HataNakaoMikhailov12}. Global, mean field term can be also accommodated for so as to induce the sought behavior. When the dynamics flow on a network,  topology matters and does play a prominent role in eliciting the instability \cite{AsllaniChallengerPavoneSacconiFanelli14, ContemoriDiPattiFanelliMiele16}. This observation motivates in the search of alternative control protocols, which leave the reaction part unchanged, while acting on the underlying web of inter-nodes connections \cite{AsllaniCarlettiFanelli16, SkardalArenas16}. The aim of this paper is to contribute along this direction, by proposing a viable topological approach which is fully rooted on first principles.   

To illustrate the proposed method we shall first operate in the framework of the Complex Ginzburg-Landau Equation (CGLE) \cite{AransonKramer02} a prototypical model for nonlinear physics, whose applications range from superconductivity, superfluidity and Bose-Einstein condensation to liquid crystals and strings in field theory.  The CGLE assumes a population of oscillators, described in terms of their complex amplitude and 
mutually coupled via a diffusive-like interaction. This is mathematically described in terms of a discrete Laplacian operator. The CGLE admits a uniform fully synchronized solution,  the spatially extended replica of the periodic orbit displayed by the system in its a-spatial version, provided the Laplacian is balanced (equal incoming and outgoing connectivity) \cite{Nakao14}. Hereafter, we shall assume that the nodes of the network where oscillators lie are initially paired (and the reaction parameters set) so as to make the system unstable to externally injected, non homogeneous, perturbations. The network of connections is then globally reshaped (keeping the reaction parameters unchanged) to regain the stability of the synchronized, time dependent, solution. We will then move forward to considering a system of coupled (real) Ginzburg-Landau equations \cite{2realGL}, which admits a stationary stable fixed point. Turing-like instabilities will be controlled, hence formally impeded, with a supervised intervention targeted to the net of interlaced couplings.     

The paper is organized as follows. In the next section we will introduce the CGLE and carry out a linear stability analysis to delineate the conditions that make the spatially extended, homogenous limit cycle solution stable. We will in particular elaborate on the remarkable differences that arise when the system involves a finite and discrete collection of interlinked oscillators, as opposed to the reference case where the population of elementary constituents is made infinite and continuous. In Section III we will provide the mathematical basis for the proposed control method. The approach will be successfully tested by operating with the CGLE and assuming a symmetric network. In Section IV, we will consider a directed, although balanced, network of couplings and extend to this setting the analysis. Finally in Section V we will sum up and draw our conclusions. The appendix \ref{app_2rGL} is devoted to discussing the stabilization of a constant homogeneous fixed point, and proving, also in this respect, the adequacy of the proposed recipe.

 \section{Complex Ginzburg-Landau equation: linear stability analysis}

Consider an ensemble made of $N$ nonlinear oscillators and label with  $W_i$ their associated complex amplitude, where $i=1,...,N$.  Each individual oscillator obeys to a CGLE which, as will be clarified in the following,
 combines linear and nonlinear (cubic) contributions. In addition, we assume the oscillators to be mutually coupled via a diffusive-like interaction which is mathematically exemplified via the discrete Laplacian operator. To fix ideas consider as an example the simplified setting in which the oscillators are coupled to nearest neighbors only,  on a one dimensional lattice complemented with periodic boundary conditions. The network of connections yields a binary adjacency matrix, termed $\mathbf{A}$: the entry $A_{ij}$ is set to one if nodes $i$ and $j$ are paired together, or zero if the link is absent.
The associated discrete Laplacian operator $\mathbf{\Delta}$  results in a circulant matrix with three non trivial entries per row, namely $\Delta_{ii}=-2$, $\Delta_{i,i+1}=1$ and $\Delta_{i,i-1}=1$. In general, a complex web of inter-nodes connections yields a heterogenous network, potentially directed ($A_{ij}\neq A_{ji}$) and weighted. 
In this case, let us denote with $k_i^{out}=\sum_jA_{ji}$ (resp. $k_i^{in}=\sum_jA_{ij}$) the outgoing (resp. ingoing) connectivity of generic node $i$. 
In our study we will deal with symmetric or balanced and directed networks, hence $k_i^{out}=k_i^{in} \equiv k_i$. 
The elements of the Laplacian matrix $\mathbf{\Delta}$ are therefore defined as $\Delta_{ij}=A_{ij}-k_i \delta_{ij}$, where $\delta_{ij}$ stands for the usual Kronecker $\delta$.  
 
The spatially extended CGLE  can be hence cast in the form: 

\begin{equation}\label{eq:GLnetwork}
\frac{d}{d t} W_j =  W_j - (1+i c_2) \vert W_j \vert ^2 W_j + (1+i c_1) K \sum_k \Delta_{jk} W_k
\end{equation}
where $c_1$ and $c_2$ are real parameters, which can be externally assigned. The index $j$ runs from $1$ to $N$, the size of the inspected system. Here, $K$ is a suitable parameter setting the coupling strength. We shall begin by considering a symmetric adjacency matrix and 
postpone to a later stage the case of a directed, though balanced, network of couplings. For pedagogical reasons, let us start by considering a regular lattice, embedded on a Euclidean space of arbitrary dimension. By performing the continuum limit, i.e assuming the linear distance between neighbor nodes to asymptotically vanish,  one can formally replace the discrete variable $W_j(t)$ ($j=1,2,\cdots, N$) with its continuous counterpart $W(\mathbf{r},t)$. Here, $W(\mathbf{r},t) \in\mathbb{C}$ and $\mathbf{r}$ identifies the space location. Under these conditions, the discrete operator $\mathbf{\Delta}$ transforms into $\nabla^2$, the standard Laplacian on a continuous support. For this reason, and with a slight abuse of language, we shall often employ the adjective spatial to tipify the nature of the coupling, even when the network of oscillators is not necessarily bound to a physical space.    
 
As a preliminary remark we note that ${W}^{LC}(t) = e^{-\mathfrak{i} c_2 t}$ is a homogenous solution of the CGLE, both in its discrete or continuous, spatially extended, versions. This latter can be referred to as the limit cycle (LC) solution, since it  results from a uniform, fully synchronized, replica of the periodic orbit displayed by the system in its a-spatial ($K=0$) version. In the remaining part of this section, we shall determine the stability of the LC solution. We will deal at first with the continuous version of the model and re-derive for completeness the conditions for the onset of the so called Benjamin-Feir instability~\cite{BenjaminFeir67, StuartDiPrima78}. The peculiarities that stem from assuming a discrete and heterogeneous web of symmetric couplings will be also reviewed. 

To assess the stability of the LC solution we introduce a non homogeneous perturbation, both in phase and amplitude: 

 \begin{equation}
\label{ansatz}
 W(\mathbf{r},t)= {W}^{LC}(t) [1+\rho(\mathbf{r},t)]e^{\mathfrak{i}\theta(\mathbf{r},t)}.
\end{equation}

Linearizing around the LC ($\rho(\mathbf{r},t)=0$, $\theta(\mathbf{r},t)=0$) one readily obtains:
 \begin{equation}
  \frac{d }{dt} \left[ \begin{array}{c} \rho \\ \theta \end{array} \right] = \begin{bmatrix} -2 & 0 \\ -2c_2 & 0 \end{bmatrix} \left[ \begin{array}{c} \rho \\ \theta \end{array} \right] + K \begin{bmatrix} 1 & -c_1 \\ c_1 & 1 \end{bmatrix} \nabla^2 \left[ \begin{array}{c} \rho \\ \theta \end{array} \right].
  \label{CGL_LinEq_cont}
 \end{equation}
 
 To solve the  above linear problem we perform a space-time Fourier transform:
 
 \begin{equation}
 \begin{array}{ll}
  \rho(\mathbf{r},t)=\int \int d\omega  d\mathbf k  e^{\mathfrak{i}\omega t}e^{\mathfrak{i}\mathbf{k}\cdot \mathbf{r}}  \rho_{\mathbf{k}} \\
  \theta(\mathbf{r},t)=\int \int d\omega  d\mathbf k  e^{\mathfrak{i}\omega t}e^{\mathfrak{i}\mathbf{k}\cdot \mathbf{r}}  \theta_{\mathbf{k}} .\\
 \end{array}  
  \label{FT}
 \end{equation}
 A straightforward calculation returns the following condition that should be matched as a necessary consistent requirement for the linear problem to admit a meaningful solution:
 
 \begin{equation}
  \label{det}
  \det \begin{bmatrix} \lambda+2+Kk^2 &\ \ \ -Kc_1k^2 \\
  \\
  2c_2+Kc_1k^2 &\ \ \ \lambda+Kk^2 \end{bmatrix} = 0
 \end{equation}
 with $\lambda=\mathfrak{i}\omega$ and $k=|\mathbf{k}|$. The quantity $\lambda$ hence assesses the linear growth rate associated to the $k$-th mode. Without losing generality we will hereafter set the coupling constant to unit ($K=1$) and proceed with the calculation to determine the root of the characteristic polynomial with largest real part: 

 \begin{equation}
\label{disprel}
  \lambda(k^2) = -k^2-1 + \sqrt{-c_1^2 k^4 - 2c_1c_2 k^2+1}.
 \end{equation}

The perturbation that shakes the homogenous and time dependent solution  ${W}^{LC}(t)$ gets exponentially magnified in the linear regime of the evolution provided the real part of  $\lambda$ (the dispersion relation $\lambda_{Re}$) is  positive. Notice that $\lambda(0)=\lambda_{Re}(0) =0$, as expected, based on a obvious argument of internal coherence. Expanding Eq.~(\ref{disprel}) for small $k$ returns
$\lambda_{Re} \simeq -(1 + c_1c_2) k^2$. The stability of the synchronized LC solution is therefore lost when $(1 + c_1c_2) <0$,   the standard condition for the onset of the Benjamin-Feir instability.  

We now turn to considering the case of a heterogenous, although symmetric, network of connections among oscillators.  To investigate the conditions that are to be met for a symmetry breaking instability of the homogeneous LC solution to develop, we proceed in analogy with the above and set  $W_j(t)=W^{LC}(t)[1+\rho_j(t)]e^{i\theta_j(t)}$, with a clear meaning of the symbols. Plugging this latter expression in 
the CGLE~(\ref{eq:GLnetwork}) and expanding to the first order in the perturbation amount, one obtains the obvious generalization of system (\ref{CGL_LinEq_cont}):

\begin{equation}
 \frac{d }{dt}  \left[ \begin{array}{c} \rho_j \\ \theta_j \end{array} \right] = \begin{bmatrix} -2 & 0 \\ -2c_2 & 0 \end{bmatrix} \left[ \begin{array}{c} \rho_j \\ \theta_j \end{array} \right] +  \begin{bmatrix} 1 & -c_1 \\ c_1 & 1 \end{bmatrix} \sum_k \Delta_{jk} \left[ \begin{array}{c} \rho_k \\ \theta_k \end{array} \right].
  \label{CGL_LinEq_net}
 \end{equation}
where we recall that $K=1$.  For regular lattices, the Fourier transform is usually invoked to solve the system of equations homologous to (\ref{CGL_LinEq_net}). This amounts to expanding the spatial perturbations on a set of planar waves, the eigenfunctions of the continuous Laplacian operator. When the system is instead defined on a network, an analogous procedure can be employed. To this end, we define the eigenvalues and eigenvectors of the discrete Laplacian operator:

\begin{equation}
 \label{eigen}
\sum_j \Delta_{ij}  \phi_j^{(\alpha)}=\Lambda^{(\alpha)} \phi_i^{(\alpha)} \quad \alpha=1,..., N
\end{equation}

When the network is undirected, the Laplacian operator is symmetric. Therefore, the eigenvalues $\Lambda^{(\alpha)}$ are real and the eigenvectors  $\boldsymbol{\phi}^{(\alpha)}$ form an orthonormal basis. This condition needs to be relaxed when dealing with the more general setting of a directed graph, as we shall discuss in the second part of the paper\footnote{A diagonalizable and connected Laplacian matrix is instead a minimal requirement to be satisfied by our analytical treatment both in the symmetric and in the directed case.}. The symmetric Laplacian matrix $\mathbf{\Delta}$ has a single zero eigenvalue
$\Lambda^{(\alpha=1)}$ corresponding to the uniform eigenvector and all other eigenvalues are negative. The indices $\alpha$ are sorted so as to satisfy
$0= \Lambda^{(1)} >\Lambda^{(2)} \ge \cdots \ge \Lambda^{(N)}$.

The inhomogeneous perturbations $\rho_j$  and $\theta_j$  can be expanded as:

 \begin{equation}
 \label{ansatz_net}
 \left[ \begin{array}{c} \rho_j \\ \theta_j \end{array} \right] = \sum_{\alpha=1}^N \left[ \begin{array}{c} \rho^{(\alpha)} \\ \theta^{(\alpha)} \end{array} \right]e^{\lambda t}\phi_j^{(\alpha)}.
 \end{equation}
 
By inserting Eq.~\eqref{ansatz_net} in Eq.~(\ref{CGL_LinEq_net}) and making use of relation~(\ref{eigen}) one eventually gets a condition formally equivalent to expression~(\ref{det}). As an important difference, the eigenvalues of the continuous Laplacian, $-k^2$, are replaced by the discrete (real and negative)  quantities  $\Lambda^{(\alpha)}$, the eigenvalues of the discrete Laplacian. Insisting on the analogy, it is of immediate evidence that the instability rises for a CGLE defined on a symmetric network when $1+c_1 c_2 <0$, a dynamical condition identical to that obtained when operating under the continuous,  by definition regular, viewpoint.  The quantity $\Lambda^{(\alpha)}$ constitutes the analogue of the wavelength for a spatial pattern in a system defined on a continuous regular lattice. It is this latter quantity which determines the spatial characteristic of the emerging patterns, when the system is defined on a heterogeneous complex support. 

\begin{figure}
   \vspace{-6cm}
   \hspace*{-2.7cm}
     \includegraphics[width=14cm]{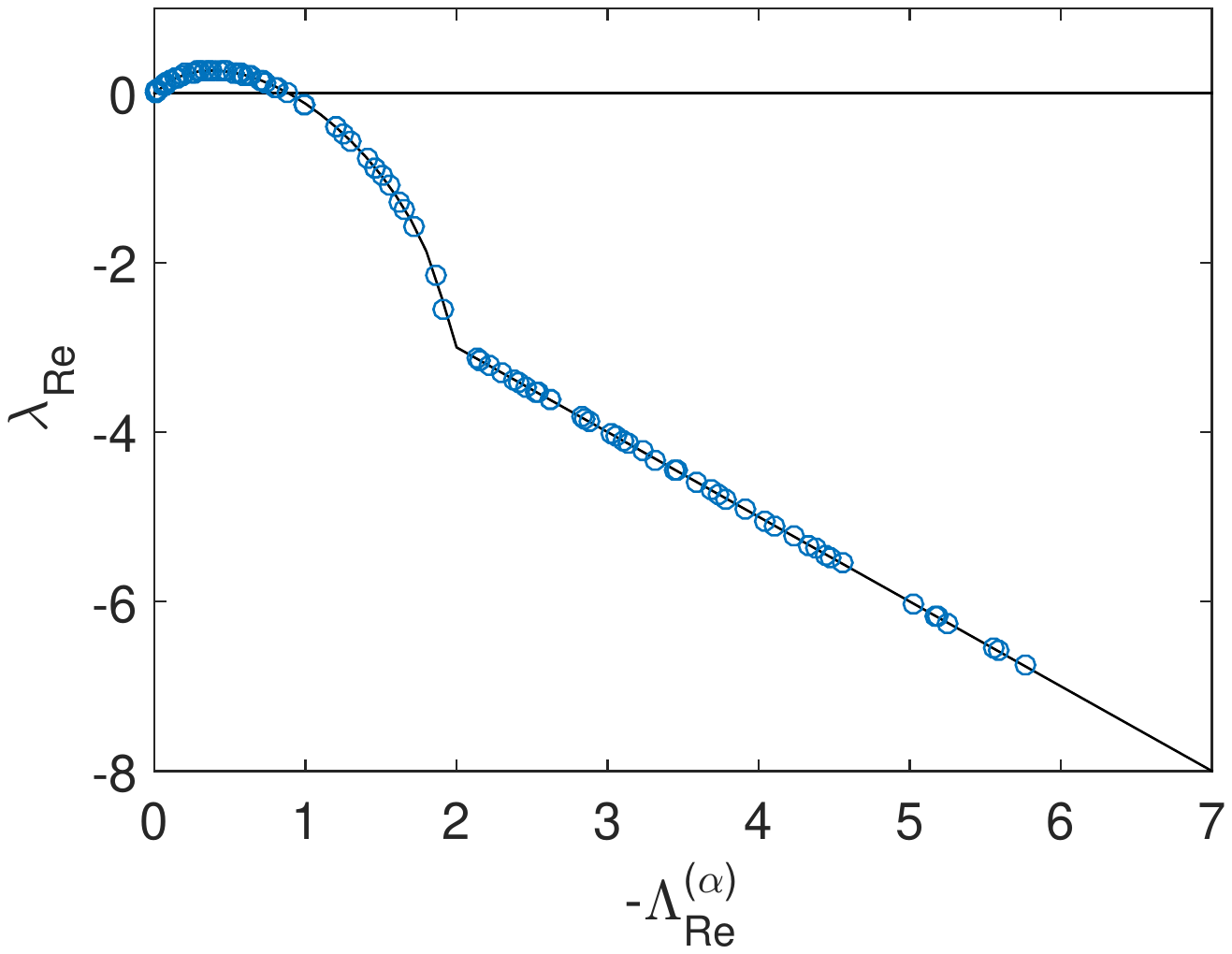}
   \vspace{-6.5cm}
   \caption{Continuous (solid line) and discrete (blue circles) dispersion relation. Here, $c_1=-1.8$, $c_2=1.6$, $K=1$. The network, composed of 100 nodes, is generated from the Watts-Strogatz method with rewiring probability 0.8.}
   \label{fig_CGL_sym_relDisp_before}
  \end{figure}

\begin{figure}
  \vspace*{-5cm}
  \hspace*{-2.3cm}
   \includegraphics[width=13.5cm]{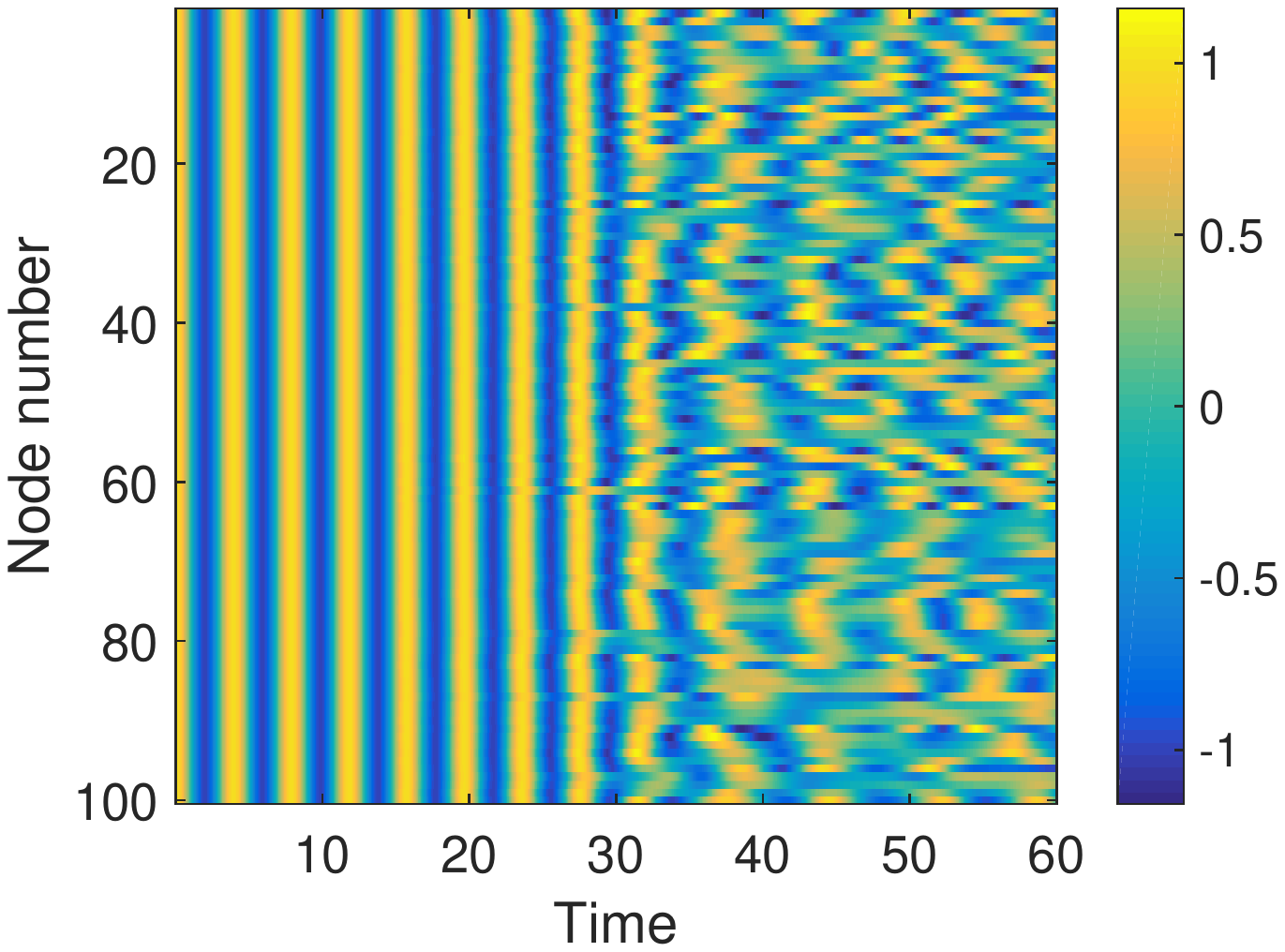}\\
   \vspace*{-12cm}
   \hspace*{-2.3cm}
   \includegraphics[width=13.5cm]{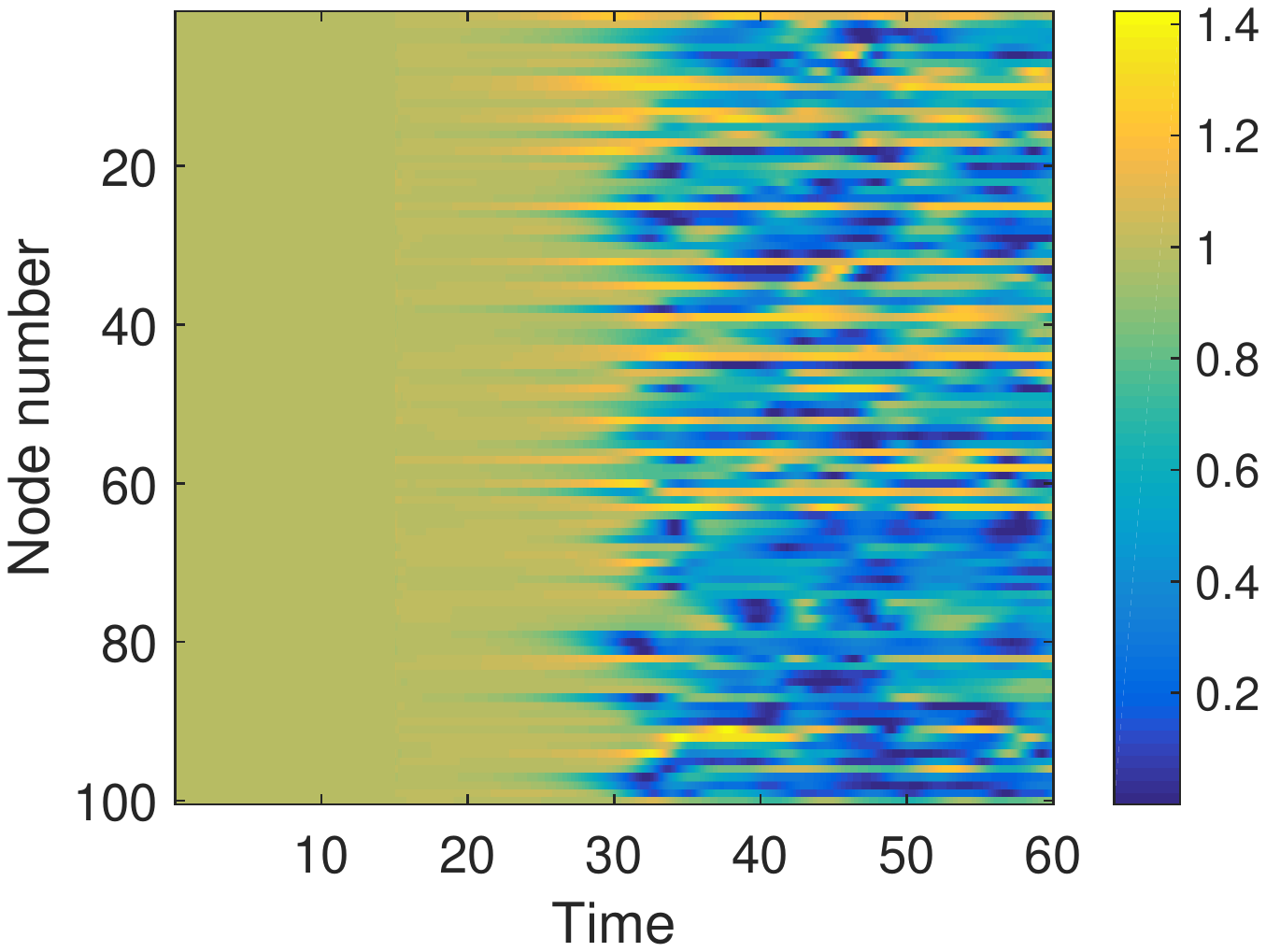}
   \vspace{-6.5cm}
   \caption{Evolution of $\mathbf W_{Re}$ (upper panel) and $|\mathbf W|^2$ (lower panel) versus time, assuming a uniform LC initial condition. At time $\tau_1=15$, a non homogeneous perturbation is inserted and the synchronized state is consequently disrupted. Here, $c_1=-1.8$, $c_2=1.6$.  The nonlinear oscillators are mutually linked via the Watts-Strogatz network, used in depicting the discrete dispersion relation of Fig.~\ref{fig_CGL_sym_relDisp_before}. 
  }
  \label{fig_CGL_sym_3behav_beginning_mod}
 \end{figure}

In Fig.~\ref{fig_CGL_sym_relDisp_before} the dispersion relation $\lambda_{Re}$ is plotted versus $-\Lambda^{(\alpha)}_{Re}$, for a specific choice of $c_1$ and $c_2$, so that $1+c_1 c_2<0$.  The solid line refers to the continuum setting ($-\Lambda^{(\alpha)}_{Re} \rightarrow k^2$,), while circles are obtained when operating with the CGLE, hosted on a Watts-Strogatz network~\cite{WattsStrogatz98}. As anticipated, the discrete collection of points which defines the dispersion relation when a symmetric, finite and heterogenous network of coupling is accommodated for, follows the same profile which applies to the limiting continuum setting. In Fig. ~\ref{fig_CGL_sym_3behav_beginning_mod} the temporal evolution of the system is displayed for a choice of the parameters which corresponds to the unstable dispersion relation of Fig.~\ref{fig_CGL_sym_relDisp_before}. After a given time the synchronized LC solution is perturbed by insertion of an external source of non homogenous disturbance. This latter grows, as predicted by the linear stability analysis, and yields the irregular patterns displayed for both $\mathbf{W}_{Re}$ and $|\mathbf W|^2$.

Back to Fig.~\ref{fig_CGL_sym_relDisp_before}, it is however important to realize that the instability actually takes place only when at least one eigenvalue $-\Lambda^{(\alpha)}_{Re}$ exists in the range where $\lambda_{Re}$  is positive. If the ensemble of discrete modes, which ultimately reflects the topology of the imposed couplings, populates the portion of the dispersion relation with $\lambda_{Re}<0$, no instability can develop, even if $1+c_1 c_2 <0$. Stated differently the  spectral gap $\Delta (\Lambda)$, the difference between the moduli of the two largest eigenvalues of the Laplacian operator ($\Delta (\Lambda)=|\Lambda^{(2)}|-|\Lambda^{(1)}|=|\Lambda^{(2)}|$) should be larger than $-2(c_1 c_2 +1)/(1+c_1^2)$, the non trivial root of Eq.~(\ref{disprel}), for the instability to take place.

This observation has been exploited by Nakao in Ref.~\cite{Nakao14} to propose a novel control strategy aimed at suppressing the Benjamin-Feir instability and thus preserving the initial synchronized regime for a CGLE defined on a symmetric network support. Imagine to start with an unstable condition, which in turn 
implies to operate with a suitable choice for both the reaction parameters and the network specificity.
The key idea of Ref.~\cite{Nakao14} is to randomly rewire the network so as to make the second eigenvalue progressively more negative. Random moves are accepted or rejected following a   
Metropolis scheme.  The numerical procedure converges to a (globally) modified network which has no eigenvalue in the range where 
$\lambda_{Re}>0$. The control is topological since it only affects the couplings that links the oscillators, without acting on the dynamical parameters $c_1$ and $c_2$. In practice, the discrete network-like system can be made stable for a choice of the parameter that would drive a Benjamin-Feir instability in the continuum limit. Building on these intriguing observations, we will here devise an analytical approach that enables us to implement a similar control protocol, without resorting to an iterative, numerically supervised, rewiring. Importantly, the method that we shall introduce here can be successfully extended to the general case where a directed network of connections is assumed to hold.  The next section is devoted to discussing the proposed method.
  
 \section{Global topological control}
 
 As outlined in the preceding section, here we aim at developing an apposite control strategy which acts on the global network of connections, leaving unchanged the dynamical parameters of the model. The method that we shall hereafter discuss takes inspiration from the seminal work of Nakao~\cite{Nakao14}. There it was shown that a numerically supervised rewiring of the inter-oscillators couplings can stabilize the CGLE, thus preserving the consensus state. Building on similar grounds, we will provide in the following an analytical procedure to achieve the sought stabilization. The proposed method allows one to immediately generate the controlled matrix of contacts, without involving any iterative scheme, thus freeing from concerns on the numerical convergence. Starting from a condition of instability, as displayed in Fig.~\ref{fig_CGL_sym_relDisp_before}, we wish to modify the spectrum of the Laplacian operator so as to force the finite and discrete collection of modes to populate the negative branch of the  dispersion relation $\lambda_{Re}$.

 As emphasized in the previous section, when the network is undirected the discrete dispersion relation superposes to the continuum one (the solid line in Fig.~\ref{fig_CGL_sym_relDisp_before}). The instability localizes on a finite set of modes, those falling on the positive bump of the curve $\lambda_{Re}(k^2)$. Is it possible to alter the network topology so as to make the (negatively defined and real) eigenvalues larger in absolute value than 
 $-2(c_1 c_2 +1)/(1+c_1^2)$,  the point where the parabola $\lambda_{Re}(k^2)$ crosses the horizontal axis, so turning negative? In a figurative sense, we want to slide the discrete points of Fig.~\ref{fig_CGL_sym_relDisp_before} onto the curve, as beads on a cord, causing them to reach  
its negative branch. To answer this question we make use of simple  linear algebra tools. 
 
Let us start by defining the $N \times N$ matrix $\boldsymbol{\Phi}$ whose columns are the eigenvectors $\boldsymbol{\phi}^{(1)}, \cdots, \mathbf{\phi}^{(N)}$
of the Laplacian operator $\mathbf{\Delta}$.  Hence $\mathbf{D} = \Phi^{-1} \mathbf{\Delta} \Phi$, where $\mathbf{D}$ is the diagonal matrix 
formed by the eigenvalues $\Lambda^{(1)},...,\Lambda^{(N)}$. We then calculate the minimal corrections $\delta \Lambda^{(\alpha)}$ ($\alpha=1, \cdots, N$) that need to be imposed to shift the eigenvalues $\Lambda^{(\alpha)}$ on the stable side of the dispersion relation. The computed corrections are then organized in a diagonal matrix $\mathbf{D}'$,  so that ${D}'_{\alpha \alpha}=\delta \Lambda^{(\alpha)}$, for $\alpha=1, \cdots, N$. As a matter of facts, and to keep the formulation general, $\delta \Lambda^{(\alpha)} \in\mathbb{C}$. For the case of a symmetric network that we are bound to explore within this Section, the quantities  $\delta \Lambda^{(\alpha)}$ are however real and negative. When the original $\Lambda^{(\alpha)}$ falls in the region of stability, the corresponding correction  $\delta \Lambda^{(\alpha)}$ is set to zero. 
 
The next step of the procedure is to perform the following transformation $\mathbf{\Delta}' = \Phi \mathbf{D'} \Phi^{-1}$
and define the controlled matrix $\mathbf{\Delta_c} = \mathbf{\Delta}+\mathbf{\Delta}'$.  By construction the eigenvalues of $\mathbf{\Delta_c}$ (with the only exception of the zero eigenvalue, $\alpha=1$) are smaller than 
 $2(c_1 c_2 +1)/(1+c_1^2)$. Assume  for a moment that $\mathbf{\Delta_c}$ can be interpreted as a Laplacian operator. Hence, the obvious conclusion is that we have generated a modified adjacency matrix $\mathbf{A_c}$, hidden inside $\mathbf{\Delta_c}$, which  should engender a negative dispersion relation (when employed in the CGLE, at fixed $c_1$ and $c_2$),  thus preserving the stability of the synchronized configuration. Before concluding in this respect, one needs to prove that $\mathbf{\Delta_c}$ is indeed a Laplacian matrix, namely that (i) the entries of the matrix are real and that (ii)
 every column of the matrix sums up to zero, $\sum_i({\Delta_c} )_{ij}=0\ \forall j$). In the following we set down to prove the above properties (for both the symmetric and asymmetric settings), before turning to provide a numerical validation of the devised control procedure. As an important complement, we will also show that symmetry and balancedness are perpetuated from  $\mathbf{\Delta}$  to $\mathbf{\Delta_c}$.

 \vspace{0.5 truecm}
 
  {\bf The elements of $\mathbf\Delta_c$ are real.}  The generic entries of $\mathbf\Delta'$ can be written as:
 \begin{equation}
  \Delta'_{il}=\sum_j\Phi_{ij}D'_{jj}(\Phi^{-1})_{jl}.
  \label{d'}
 \end{equation}
 In the symmetric case, the eigenvalues and their corresponding corrections are real. Also the eigenvectors have real entries (vectors in $\mathbb{R}^N$). Hence, the elements of $\mathbf D'$, $\mathbf\Phi$, $\mathbf\Phi^{-1}$ are real and, consequently, $\Delta'_{ij} \in \mathbb{R}$.
  The undirected case is more complicated to handle. Let us bring into evidence the real and imaginary parts of every element of Eq.~(\ref{d'}). It is immediate to see that the imaginary part of $\Delta'_{il}$ reads:
 \begin{equation}
  \begin{split}
  &(\Delta'_{Im})_{il}=\\
  &=\sum_j(D'_{Im})_{jj}[(\Phi_{Re})_{ij}(\Phi^{-1}_{Re})_{jl}-(\Phi_{Im})_{ij}(\Phi^{-1}_{Im})_{jl}] +\\ 
  &+\sum_j (D'_{Re})_{jj}[(\Phi_{Re})_{ij}(\Phi^{-1}_{Im})_{jl}+(\Phi_{Im})_{ij}(\Phi^{-1}_{Re})_{jl}].
  \end{split}
 \label{ImDelta}
 \end{equation}
  
To match condition $(\Delta'_{Im})_{il}=0$, both terms on the right hand side of Eq.~(\ref{ImDelta}) should be zero. To prove this fact let us begin by recalling that the eigenvalues of a real asymmetric matrix either are real or are complex and come in conjugate pairs. Consider first the latter case and label with $\alpha$ and $\beta$ the generic pair of conjugate eigenvalues. By definition  $\mathbf\Delta\boldsymbol{\phi}^{(\alpha)}=\Lambda^{(\alpha)}\boldsymbol\phi^{(\alpha)}$. Taking the complex conjugate yields 
 $\mathbf\Delta (\boldsymbol{\phi}^{(\alpha)})^*=(\Lambda^{(\alpha)})^* (\boldsymbol\phi^{(\alpha)})^*$ where $(\cdot)^*$ stands for the complex conjugate and where use has been made of the condition  $\mathbf\Delta =  \mathbf\Delta^*$. Recalling that  $(\Lambda^{(\alpha)})^*=\Lambda^{(\beta)}$ we can immediately conclude that $(\boldsymbol{\phi}^{(\alpha)})^*$ is an eigenvector of $\mathbf\Delta$ relative to the eigenvalue $\Lambda^{(\beta)}$ and thus 
 $\boldsymbol\phi^{(\beta)}=(\boldsymbol\phi^{(\alpha)})^*$. Hence
 
 \begin{equation}
 \begin{array}{ll}
  (\Phi_{Re})_{i\alpha}=(\Phi_{Re})_{i\beta}\\
   (\Phi_{Im})_{i\alpha}=-(\Phi_{Im})_{i\beta}
 \end{array}
  \label{ReImPhi}
 \end{equation}
 for every $i$ and $(\alpha,\beta)$. Consider now the equation
 \begin{equation}
  (\boldsymbol\phi^{-1})^{(\alpha)}\mathbf\Delta=\Lambda^{(\alpha)}(\boldsymbol\phi^{-1})^{(\alpha)}
 \end{equation}
 with $(\boldsymbol\phi^{-1})^{(\alpha)}$ $\alpha$-th row of $\boldsymbol\Phi^{-1}$. Proceeding in analogy with the above, one eventually gets:
 \begin{equation}
 \begin{array}{ll}
   (\Phi^{-1}_{Re})_{\alpha l}=(\Phi^{-1}_{Re})_{\beta l}\\
   (\Phi^{-1}_{Im})_{\alpha l}=-(\Phi^{-1}_{Im})_{\beta l}.
 \end{array}
  \label{ReImPhi_1}
 \end{equation}
 Let us go back to Eq.~(\ref{ImDelta}). Performing the summation on $j=\alpha$ and $j=\beta$, using Eq.~(\ref{ReImPhi}) and Eq.~(\ref{ReImPhi_1}) and the fact that the corrections $D'_{\alpha\alpha}$ and $D'_{\beta\beta}$ are complex conjugated as the original eigenvalues  $\Lambda^{(\alpha)}$ and $\Lambda^{(\beta)}$ are, we finally conclude that the terms of the sums in Eq.~(\ref{ImDelta}) cancel in pairs. 

Consider now the case of a real eigenvalue 
$\Lambda^{(k)}$. Hence, by definition,  $(D'_{Im})_{kk}=0$, since, in this case, the stabilization can be solely achieved by acting on the real part of the eigenvalue (see next Section). To prove that $(\Delta_{Im})_{il}=0$ we need therefore to focus on the second term of Eq.~(\ref{ImDelta}), with $j=k$.  Without losing generality  (up to a constant scaling factor) $\boldsymbol\phi^{(k)}_{Im}=0$:  the eigenvector associated to  $\Lambda^{(k)}$ is hence real. The ${ik}$ entries of matrix $\boldsymbol\Phi$ are indeed the elements of $\boldsymbol\phi^{(k)}$ and, for this reason, $(\Phi_{Im})_{ik}=0\ \forall i$. To conclude the reasoning and eventually prove that $(\Delta_{Im}')_{il}=0$, one needs to show that $(\Phi_{Re})_{ik}  (\Phi^{-1}_{Im})_{kl} =0$. This is in fact the case: $(\boldsymbol\phi^{-1})^{(k)}$ is the left eigenvector of matrix $\mathbf{\Delta}$, relative to the real eigenvalue  $\Lambda^{(k)}$. Reasoning as above, 
 one can take $(\boldsymbol\phi^{-1})^{(k)}$ to be real and thus $(\Phi^{-1}_{Im})_{kl}=0$. Then, summing up, $(\Delta_{Im}')_{il}=0\ \forall i,l$.
 
\vspace{0.5 truecm}
 
  {\bf Each column of $\mathbf\Delta_c$ sums up to zero.} Consider
  \begin{equation}
  \begin{split}
  \sum_i\Delta'_{il}&=\sum_{ij}\Phi_{ij}D'_{jj}(\Phi^{-1})_{jl}=\\
  &=\sum_{j}(\Phi^{-1})_{jl}D'_{jj}\sum_i\Phi_{ij}.
  \end{split}
  \label{dim0}
 \end{equation}
 Observe that
 \begin{equation}
 \begin{split}
  \Lambda^{(\alpha)}\phi_i^{(\alpha)}&=\sum_j\Delta_{ij}\phi_j^{(\alpha)}=\sum_j(A_{ij}-\delta_{ij}k_j)\phi_j^{(\alpha)}=\\
  &=\sum_jA_{ij}\phi_j^{(\alpha)}-k_i\phi_i^{(\alpha)}=\\
  &=\sum_jA_{ij}\phi_j^{(\alpha)}-\sum_lA_{li}\phi_i^{(\alpha)}
  \end{split}
 \end{equation}
 Summing over $i$ one obtains:
 \begin{equation}
 \Lambda^{(\alpha)}\sum_i\phi_i^{(\alpha)}=\sum_{ij}A_{ij}\phi_j^{(\alpha)}-\sum_{li}A_{li}\phi_i^{(\alpha)}=0, 
 \end{equation}
 thus the sum of elements of each Laplacian's eigenvector (corresponding to an eigenvalue different from zero) is identically equal to zero. 
 This observation can be used to conclude the proof. In fact,  in Eq.~(\ref{dim0}), $\sum_i\Phi_{ij}$ is equal to zero for all $j$ associated to $\Lambda^{(j)}\neq0$. On the other hand, when $j$ corresponds to $\Lambda^{(j)}=0$, $D'_{jj}=0$, as no correction is in this limiting case needed. 
 Hence,  $\sum_i(\Delta')_{il}=0$ $\forall l$ which in turn implies $\sum_i(\Delta_{c})_{il}=0$ $\forall l$.
 
 \vspace{0.5 truecm}

 {\bf If $\mathbf\Delta$ is symmetric, then also $\mathbf\Delta_{c}$ is.} It is enough to prove that the corrections  $\mathbf\Delta'$ are bound to be symmetric. Indeed, when $\mathbf\Delta$  is symmetric, matrix $\mathbf\Phi$ is orthogonal \footnote{The eigenvectors of a symmetric matrix generate an orthonormal basis.} ($\mathbf\Phi^{-1}=\mathbf\Phi^T$). Hence:
  \begin{equation}
  \begin{split}
  \Delta'_{il}&=\sum_{j}\Phi_{ij}D'_{jj}(\Phi^{-1})_{jl}=\\
  &=\sum_{j}\Phi_{ij}D'_{jj}\Phi_{lj}=\\
  &=\sum_{j}\Phi_{lj}D'_{jj}(\Phi^{-1})_{ji}=\Delta'_{li}
  \end{split}
 \end{equation}  
which ends the proof.  
  
  \vspace{0.5 truecm}
  
{\bf If $\mathbf\Delta$ is balanced, then also $\mathbf\Delta_{c}$ is.}
 The Laplacian is termed balanced when, for every node, the ingoing connectivity equals the outgoing one ($k_i^{in}=k_i^{out}$ $\forall i$), i.e., if and only if each Laplacian row sums up to zero \footnote{The network is balanced if and only if $\sum_j\Delta_{ij}=0$. Indeed we have:
$\sum_j\Delta_{ij}=\sum_jA_{ij}-k_i^{out}=k_i^{in}-k_i^{out}=0.$
 So $\sum_j\Delta_{ij}=0$ if and only if $k_i^{in}=k_i^{out}$. Note that in the above we assumed the Laplacian defined as $\Delta_{ij}=A_{ij}-k_i^{out} \delta_{ij}$.}. It is then sufficient to prove $\sum_l\Delta'_{il}=0$.
  First of all, recall that the columns of matrix $\mathbf\Phi$ are the right eigenvectors of $\mathbf\Delta$, while the rows of $\mathbf\Phi^{-1}$ are the left eigenvectors, namely:
  \begin{equation}
   \begin{array}{ll}
    \mathbf\Delta \mathbf\Phi=\mathbf\Phi \mathbf D\\
    \mathbf\Phi^{-1}\mathbf\Delta=\mathbf D \mathbf\Phi^{-1}
   \end{array}
  \end{equation}
 By definition of Laplacian, the uniform vector $\mathbf1$ is the left eigenvector of $\mathbf\Delta$ corresponding to $\Lambda^{(1)}=0$:
 \begin{equation}
  \mathbf1^{T}\mathbf\Delta=0\ \ \Rightarrow \ \ \sum_i\Delta_{ij}=0.
 \end{equation}
 If $\mathbf\Delta$ is balanced then also the right eigenvector corresponding to $\Lambda^{(1)}=0$ is equal to $\mathbf1$, hence:
 \begin{equation}
  \mathbf\Delta \mathbf1=0\ \ \Rightarrow \ \ \sum_j\Delta_{ij}=0.
 \end{equation}
 By controlling the network of connections we modify the eigenvalues of the Laplacian operator, while keeping the eigenvectors unchanged. As a consequence,  vectors $\mathbf1$ and  $\mathbf1^{T}$ are still solutions of the right and left eigenvalue problems. Moreover, the modified Laplacian 
 $\mathbf\Delta_c$ has still zero among the eigenvalues, thus:
 
  \begin{equation}
  \begin{array}{ll}
   \mathbf{\Delta}_c\mathbf1=0\ \ \Rightarrow \ \ \sum_j(\Delta_c)_{ij}=0\\
   \mathbf1^{T}\mathbf\Delta_c=0\ \ \Rightarrow \ \ \sum_i(\Delta_c)_{ij}=0
  \end{array}
 \end{equation}
 which proves the claim.

 \vspace{0.5 truecm}

\begin{figure}
   \vspace{-6cm}
   \hspace*{-2.7cm}
   \includegraphics[width=14cm]{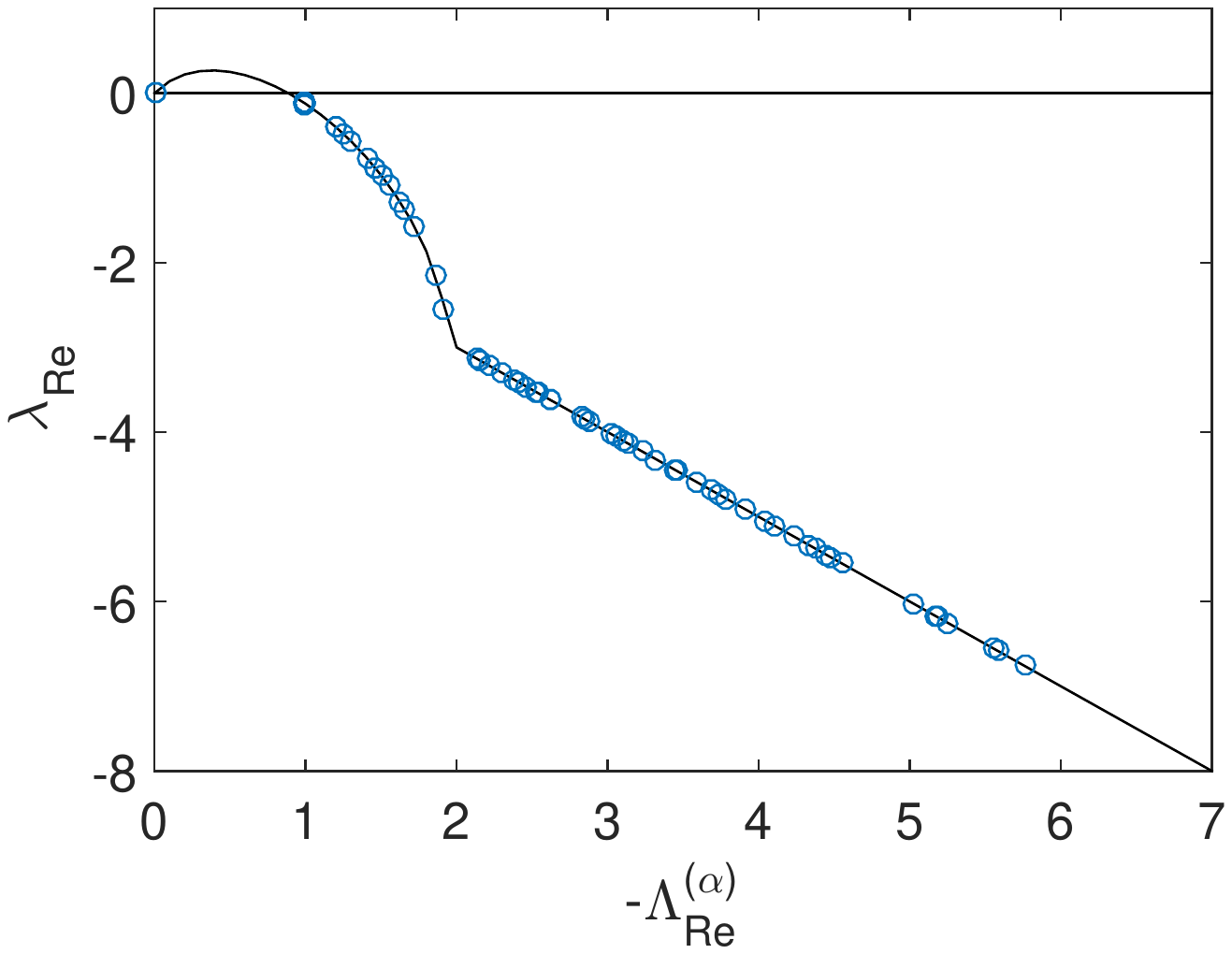}
   \vspace{-6.5cm}
   \caption{Circles show the dispersion relation obtained for the controlled adjacency matrix. The solid line refers to the dispersion relation in the continuum limit. Parameters are set as in Fig.~\ref{fig_CGL_sym_relDisp_before}.}
   \label{fig_CGL_sym_relDisp_after}
  \end{figure}

\begin{figure}
      \includegraphics[width=9cm]{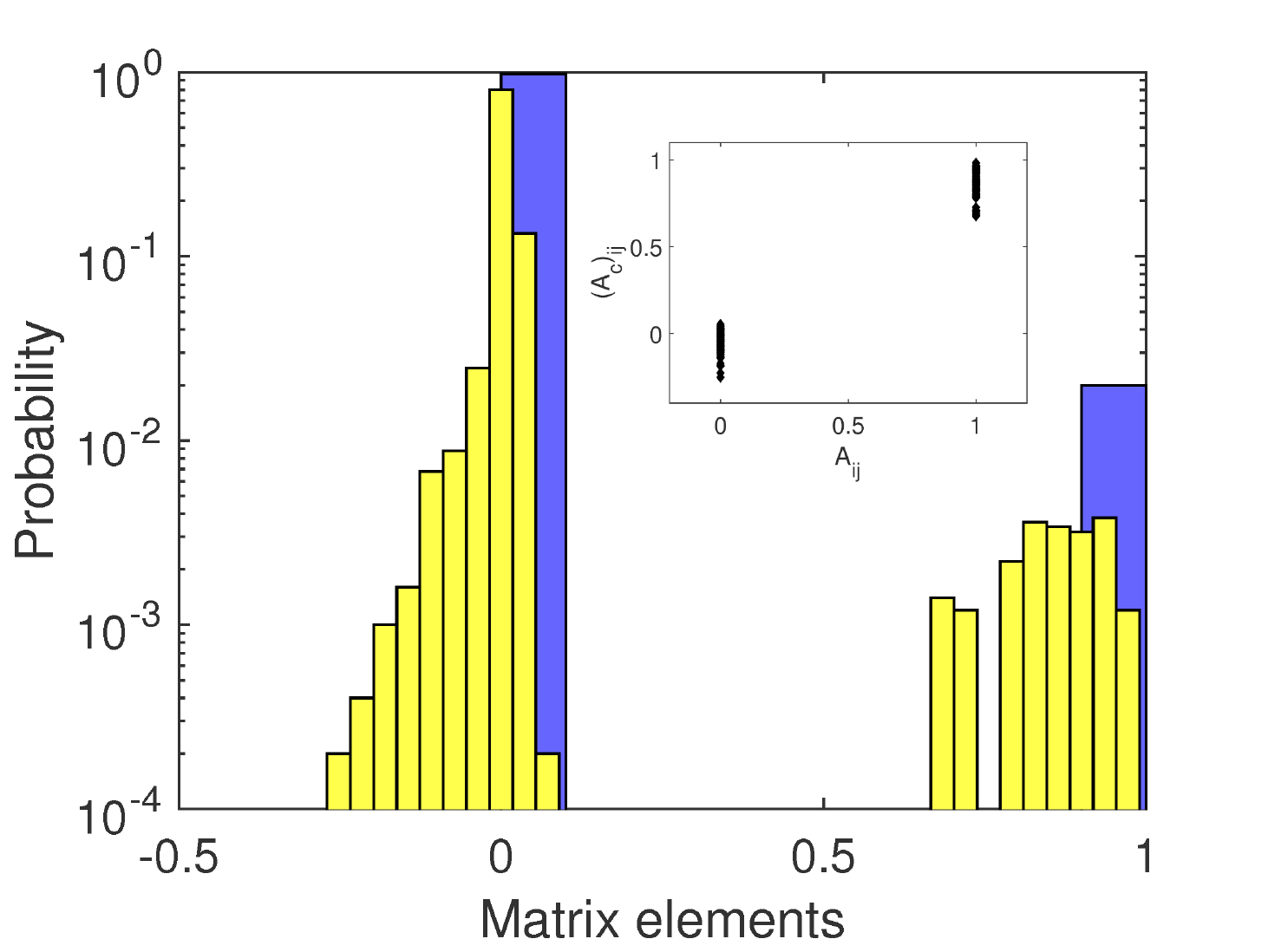}
    \caption{Main panel: distribution of the elements of the adjacency matrix before (large dark bins) and after (light small bins) the control. Initially the distribution displays two peaks localized in $0$ and $1$, reflecting the choice of a binary matrix of contacts. The controlled adjacency matrix is still bimodal, but the peaks are now smoothed out. Importantly, negative connections, pointing to inhibitory loops, should be accommodated for in the rewired weighed network. Inset: the elements of the controlled adjacency matrix $(A_c)_{ij}$ are plotted vs. the original adjacency matrix  $A_{ij}$ and a clear correlation is displayed. The control manifests as a rather local modification of the weights, strong (resp. weak) couplings being preserved under the imposed rewiring.  
    }
    \label{isto_A_sim_log}
   \end{figure}

In the remaining part of this section we will test the proposed control scheme assuming a symmetric matrix of inter-nodes couplings. In the next Section we will turn to discussing the more general case of a directed, although balanced, adjacency matrix. To demonstrate the adequacy of the technique, we will assume the setting depicted in Fig.~\ref{fig_CGL_sym_relDisp_before}: the parameters ($c_1$, $c_2$) and the underlying network of contact are chosen so as to make the system unstable to external non homogeneous perturbations. By rewiring the network following the strategy outlined above we obtain the dispersion relation represented in Fig.~\ref{fig_CGL_sym_relDisp_after}. The circles stand for the discrete dispersion relation and  populate the negative portion of the continuous curve: the instability has been hence removed, by solely acting on the topology of the graph. This latter was initially assumed of the binary type: the entries of the adjacency matrix are therefore a collection of zeros and ones. The elements of the controlled matrix are still characterized by a bimodal distribution, as displayed in Fig.~\ref{isto_A_sim_log}. 
Each element of the controlled adjacency matrix $(A_c)_{ij}$ takes a value close to the initial entry $A_{ij}$. In practice, the control returns a local adjustment of the weights, strong (resp. weak) couplings being preserved under the imposed rewiring.  Interestingly, negative coupling constants appear as a result of the continuous smoothing of the peak initially localized in zero. Inhibitory interactions should be hence at play for an effective stabilization of the dynamics. 

\begin{figure}
    \vspace*{-5cm}
    \hspace*{-2.3cm}
     \includegraphics[width=13.5cm]{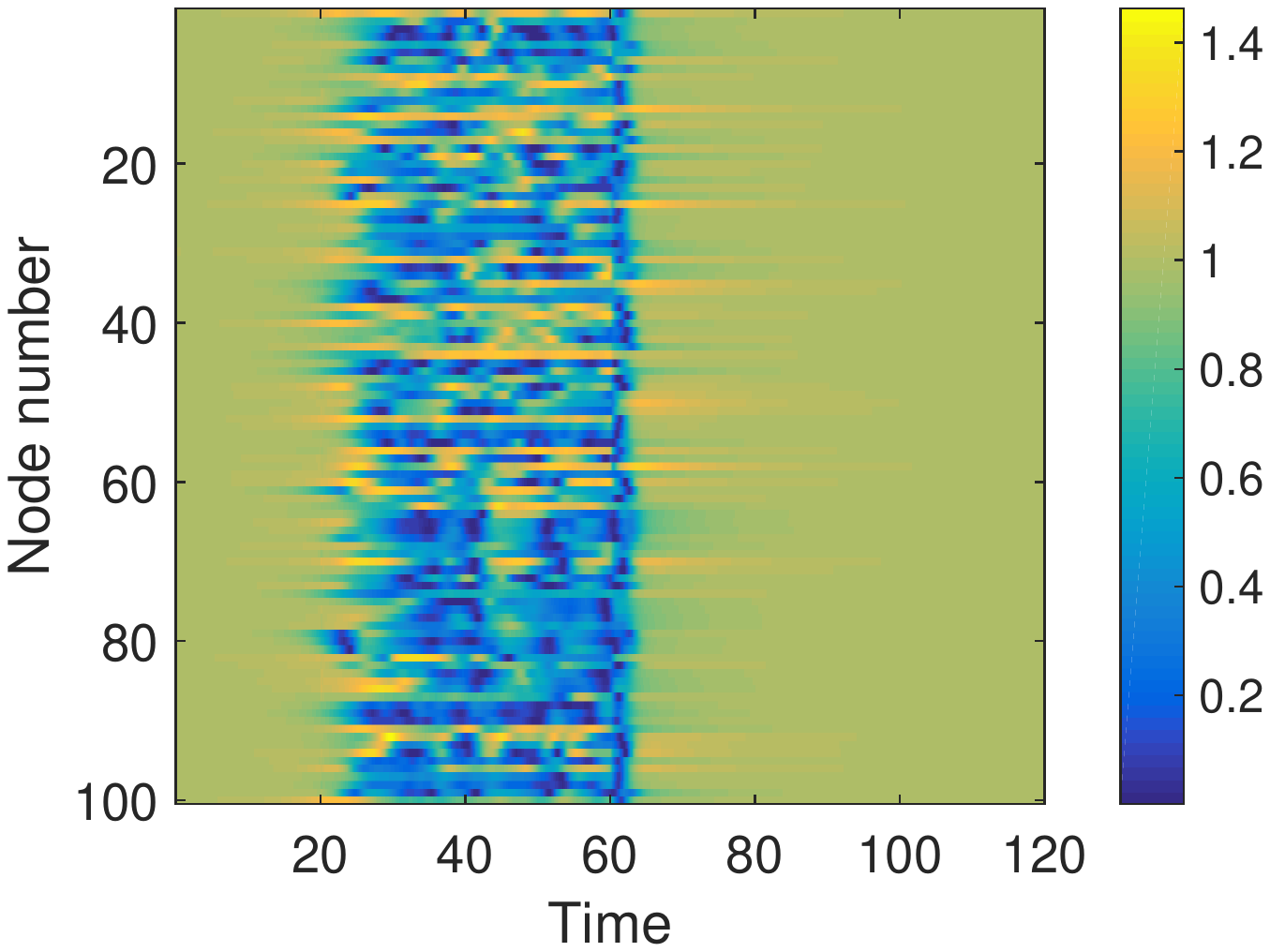}
     \vspace{-6cm}
   \caption{$|\mathbf W|^2$ vs. time. The system assumes initially a binary matrix of connections and it is unstable to external non homogenous perturbation. At time $\tau_1=10$  the LC is perturbed, and the injected disturbances grow, yielding the expected loss of synchronization. At time $\tau_2=60$ the adjacency matrix is instantaneously controlled, according to the scheme explained in the main body of the paper. The perturbation is then re-absorbed and the consensus state recovered.}
   \label{fig_CGL_sym_3behav_pert_mod}
  \end{figure}

To provide a numerical validation of our conclusion, we evolved for a transient the CGLE assuming the original, unstable and binary, adjacency matrix. When the imposed perturbation has grown to become significant, we instantaneously switched to the controlled Laplacian. As shown in Fig.~\ref{fig_CGL_sym_3behav_pert_mod}, the perturbation fades progressively away and the synchronous dynamics is eventually restored. 
The proposed control scheme was originally devised to contrast the onset of instability and, as such, targeted to the linear regime of the evolution. As demonstrated in Fig.~\ref{fig_CGL_sym_3behav_pert_mod}, the method proves however effective in stabilizing the system also at relatively large time, when nonlinearities are at play.

  \section{Controlling the instability on balanced directed networks}

 Let us now turn to considering the case of a  CGLE defined on a directed, heterogeneous although balanced (for each node the sum of incoming weights coincides with the sum of outgoing weights), network.  Before discussing the application of the control technique introduced in the previous section, we will review the conditions that determine the emergence of instability. 

 When reaction-diffusion systems are placed on directed, hence asymmetric graphs, patterns can develop, even if they are formally impeded on a symmetric, continuum or discrete, spatial support. Directionality matters and proves indeed fundamental  in shaping the emerging patterns. The conditions for the asymmetry driven instability, reminiscent of a Turing like mechanism, for a multi-species reaction diffusion model evolving on a directed graph have been discussed in Ref.~\cite{AsllaniChallengerPavoneSacconiFanelli14}. In this latter case, the perturbation acts on a homogeneous fixed point,  a time independent equilibrium for the reaction dynamics. In Ref.~\cite{DiPattiFanelliMieleCarletti16} the analysis has been extended to the setting where the unperturbed homogeneous solution is a LC and thus depends explicitly on time. In the following, for the sake of consistency, we will go through the analysis of Ref.~\cite{DiPattiFanelliMieleCarletti16} to eventually obtain the conditions that instigate the topological instability of a time-dependent solution of the LC type. 

 By perturbing the $\mathbf W^{LC}(t)$ as discussed in the first Section, one eventually ends up with the self-consistent condition:
 
 \begin{equation}
  \label{eq:disp_rel}
  \det \begin{bmatrix} -2+\Lambda^{(\alpha)}-\lambda & -c_1\Lambda^{(\alpha)} \\ -2c_2+c_1\Lambda^{(\alpha)} & \Lambda^{(\alpha)}-\lambda \end{bmatrix} = 0
 \end{equation}
  which is equivalent to $\text{det}  \left (  {\bf J}_{\alpha}  - \lambda \mathbb{I}_{2}  \right  ) =0$ with:

 \begin{equation*}
 {\bf J}_{\alpha} = \left (
\begin{matrix}
-2 +\Lambda^{(\alpha)} & -c_1 \Lambda^{(\alpha)}\\
-2 c_2 + c_1 \Lambda^{(\alpha)} & \Lambda^{(\alpha)}
\end{matrix}
\right ) \qquad .
\end{equation*}

Recall that for an asymmetric network, the Laplacian eigenvalues $\Lambda^{(\alpha)}$ are complex. Furthermore,  $\Lambda_{Re}^{(\alpha)} <0$, since the spectrum of the Laplacian matrix  falls in the left half of the complex plane, according to the Gerschgorin
theorem~\cite{Bell65}. Simple calculations yield:

\begin{equation}\label{eq:det_trace}
\begin{aligned}
(\text{tr} {\bf J}_{\alpha})_{Re}  & =  -2 + 2 \Lambda_{Re}^{(\alpha)}\\
(\text{tr} {\bf J}_{\alpha})_{Im} & =   2 \Lambda_{Im}^{(\alpha)} \\
(\text{det} {\bf J}_{\alpha})_{Re}  & =  - 2 \Lambda_{Re}^{(\alpha)} +  (\Lambda_{Re}^{(\alpha)})^2 -   (\Lambda_{Im}^{(\alpha)})^2 \\
&\phantom{ = } -2 c_1 c_2   \Lambda_{Re}^{(\alpha)} + c_1^2 \left [ (\Lambda_{Re}^{(\alpha)})^2 -  (\Lambda_{Im}^{(\alpha)})^2 \right  ]\\
(\text{det} {\bf J}_{\alpha})_{Im}  & =  -2   \Lambda_{Im}^{(\alpha)} + 2 (1 + c_1^2)\Lambda_{Re}^{(\alpha)} \Lambda_{Im}^{(\alpha)} \\
&\phantom{ = } - 2 c_1 c_2 \Lambda_{Im}^{(\alpha)}  \Lambda_{Im}^{(\alpha)}
\end{aligned}
\end{equation}

with a clear meaning of the chosen notation.

From Eq.~(\ref{eq:disp_rel}), one gets:
 
\begin{equation}
\lambda=\frac{1}{2}\left[(\text{tr} {\bf J}_{\alpha})_{Re}+\gamma \right]+\frac{1}{2}\left[(\text{tr} {\bf J}_{\alpha})_{Im}+\delta \right] \mathfrak i
\end{equation}

where:

\begin{equation}
\gamma=\sqrt{\frac{a+\sqrt{a^2+b^2}}{2}}
\end{equation}

\begin{equation}
\delta=sgn(b) \sqrt{\frac{-a+\sqrt{a^2+b^2}}{2}}
\end{equation}
and:

\begin{equation}\label{eq:a_b}
\begin{aligned}
a &=[(\text{tr} {\bf J}_{\alpha})_{Re}]^2 - [(\text{tr} {\bf J}_{\alpha})_{Im}]^2 - 4 (\text{det} {\bf J}_{\alpha})_{Re}\\
b &= 2 (\text{tr} {\bf J}_{\alpha})_{Re} (\text{tr} {\bf J}_{\alpha})_{Im} - 4 (\text{det} {\bf J}_{\alpha})_{Im}.
\end{aligned}
\end{equation}
As discussed in Ref.~\cite{AsllaniChallengerPavoneSacconiFanelli14, DiPattiFanelliMieleCarletti16},  diffusion driven instabilities arise also when $\text{tr}({\bf J}_{\alpha})_{Re}<0$, as opposed to what it happens when the system evolves on a symmetric spatial support. In fact, $\lambda_{Re}>0$ if:

\begin{equation}\label{eq:inequality}
\vert (\text{tr} {\bf J}_{\alpha})_{Re}  \vert \leqslant  \sqrt{\frac{a+\sqrt{a^2+b^2}}{2}}
\end{equation}
 a condition that can be met for $\text{tr} ({\bf J}_{\alpha})_{Re}<0$, if the network of interactions is made directed and, consequently, an imaginary component of the Laplacian spectrum is accommodated for. A straightforward, though lengthy, calculation  allows one to derive the following compact condition for the topology instability to occurr:

 \begin{equation}
 \label{eq:stabilityReteGen}
 S_2(\Lambda_{Re}^{(\alpha)}) \leq S_1(\Lambda_{Re}^{(\alpha)}) \left [ \Lambda_{Im}^{(\alpha)} \right ] ^2
 \end{equation}
 where 
 \begin{equation}
 \begin{split}
  S_2(\Lambda_{Re}^{\alpha}) &=C_{2,4}(\Lambda_{Re}^{(\alpha)})^4-C_{2,3}(\Lambda_{Re}^{(\alpha)})^3+C_{2,2}(\Lambda_{Re}^{(\alpha)})^2\\
  & \phantom{= }-C_{2,1}\Lambda_{Re}^{(\alpha)} \\
  S_1(\Lambda_{Re}^{\alpha}) &=C_{1,2}(\Lambda_{Re}^{\alpha})^2 -C_{1,1}\Lambda_{Re}^{(\alpha)} +C_{1,0}
 \end{split}
 \end{equation}
with
\begin{equation}
\label{eq:coeff}
\begin{split}
C_{2,4}&=1+c_{1}^2  \\
C_{2,3}&=4+2c_{1}c_{2}+2c_{1}^2 \\
C_{2,2}&=5+4c_{1}c_2+c_{1}^2 \\
C_{2,1}&=2+2c_{1}c_2 \\
C_{1,2}&=c_1^4+c_1^2 \\
C_{1,1}&=2c_1^{3}c_2+2c_1^2  \\
C_{1,0}&=c_{1}^{2}(1+c_2^2) \qquad .
\end{split}
\end{equation}

Notice that Eq.~(\ref{eq:stabilityReteGen}) reduces to $S_2(\Lambda_{Re}^{(\alpha)})\leq0$ when dropping the imaginary components of  $\Lambda^{(\alpha)}$, or, equivalently, when 
assuming a symmetric network of couplings. Expanding the solution for small $\Lambda_{Re}^{(\alpha)}$, assumed as a continuum variable, one readily gets $1+c_1 c_2 <0$, i.e. the standard condition for the Benjamin-Feir instability on a symmetric support.

To gain insight into the above analysis, we generate a directed and balanced network, via a suitable modification of the Newman-Watts (NW) algorithm~\cite{NewmanWatts99}. We begin from a substrate $L$-regular ring made of $N$ nodes and add, on average, $NLp$ long-range directed links. Here, $p \in [0, 1]$ is a probability that quantifies the amount of introduced  long-range links.  To keep the network balanced,  the insertion of a long-range link stemming from node $i$ is followed by a fixed number ($3$ is our arbitrary choice) of additional links to form a loop that closes on $i$~\cite{AsllaniChallengerPavoneSacconiFanelli14}.  

\begin{figure}
   \vspace{-5cm}
   \hspace*{-2.7cm}
    \includegraphics[width=14cm]{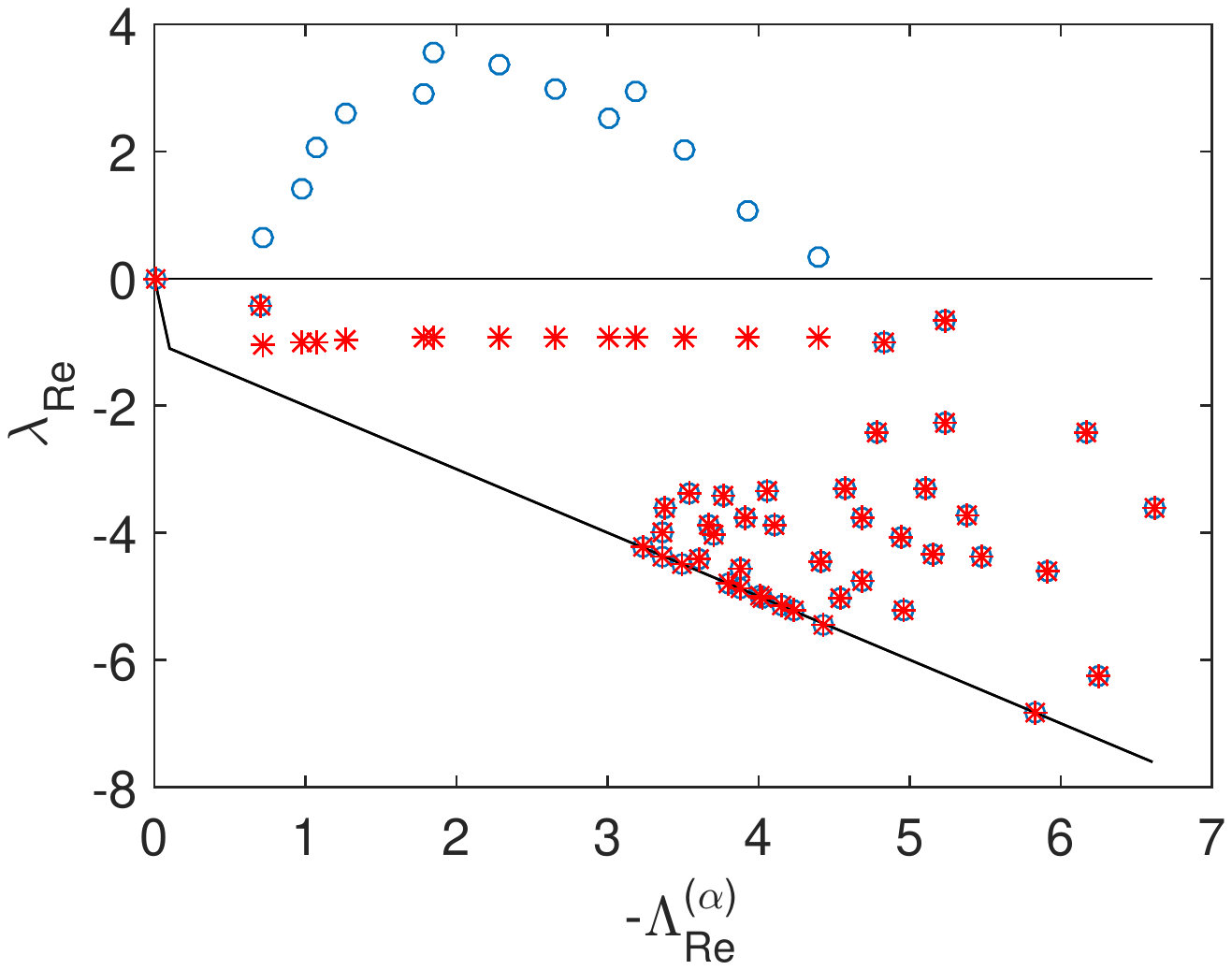}
   \vspace{-6.5cm}
   \caption{The dispersion relation $\lambda_{Re}$ as a function of $-\Lambda_{Re}^{(\alpha)}$. The solid line stands for the continuum dispersion relation. The (blue online) circles are obtained for the CGLE defined on a NW balanced network, with $p=0.27$. The (red online) stars represent the dispersion relation obtained for the controlled matrix of couplings. Here, $c_1=3$ and $c_2=2.4224$.}
   \label{fig_disp_rel_directed}
  \end{figure}

In Fig.~\ref{fig_disp_rel_directed} the dispersion relation $\lambda_{Re}$ is plotted as a function of $-\Lambda_{Re}^{(\alpha)}$. 
The black solid line refers to the limiting case of a symmetric (and continuum) support: the reaction parameters $(c_1,c_2)$ are chosen so as to prevent the instability to develop since $\lambda_{Re}<0$. The (blue online) circles refer instead to the directed case: the points abandon the solid curve and lift above zero, signaling a topology driven instability of the uniform LC solution.

\begin{figure}
   \vspace{-5cm}
   \hspace*{-3cm}
     \includegraphics[width=14cm]{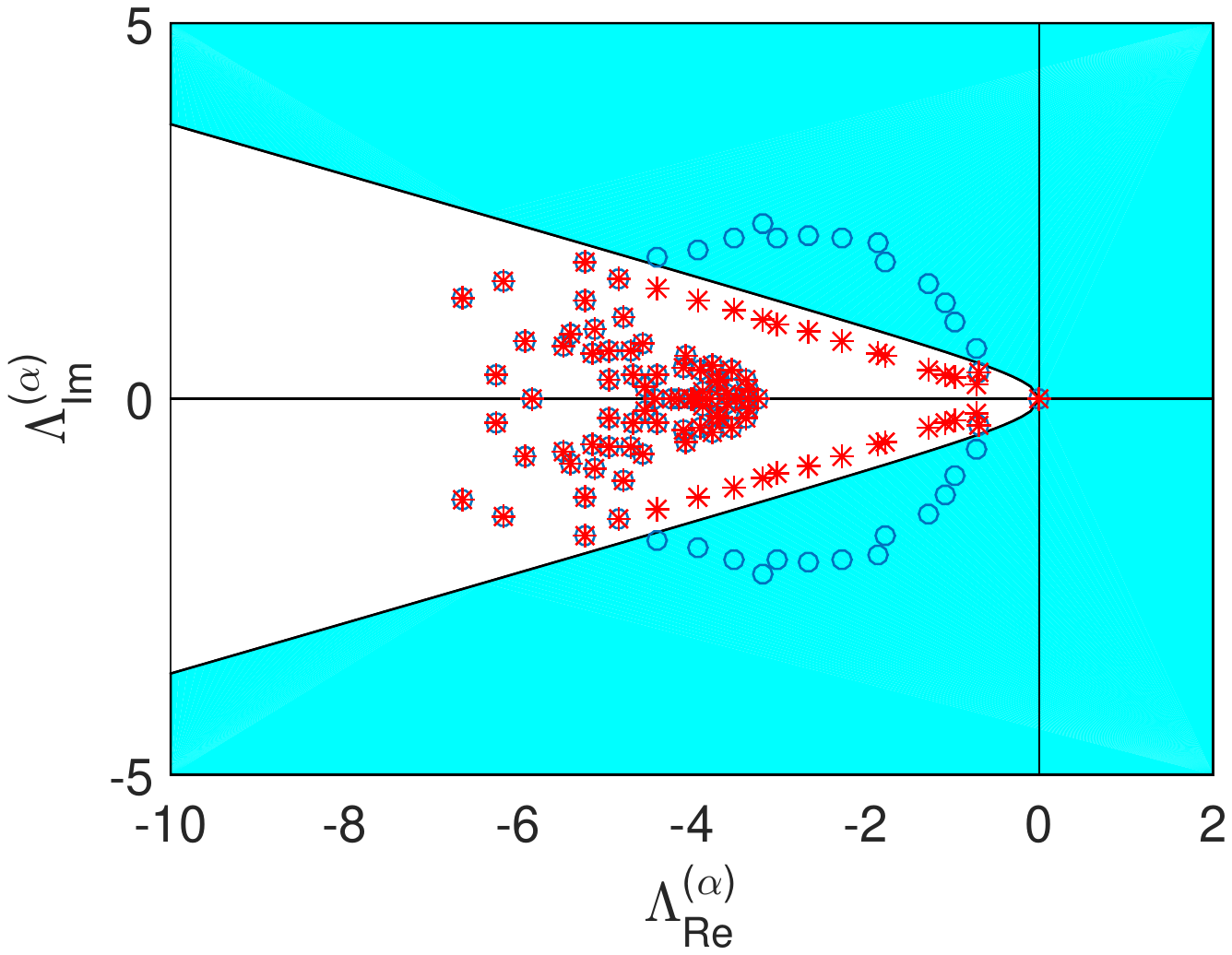}
   \vspace{-6.5cm}
   \caption{Eigenvalues in the complex plane $(\Lambda_{Re}^{(\alpha)},\Lambda_{Im}^{(\alpha)})$. The blue circles represent the eigenvalues of the initial Laplacian $\mathbf\Delta$, while the red stars are the eigenvalues of $\mathbf\Delta_{c}$. The shaded area represents the instability region obtained from Eq.~(\ref{eq:stabilityReteGen}).  The eigenvalues in this region correspond to the unstable modes, characterized by  $\lambda_{Re}>0$,  in Fig.~\ref{fig_disp_rel_directed}.}
   \label{fig_CGLstabReg_after}
  \end{figure}

In Fig.~\ref{fig_CGLstabReg_after} the same situation is illustrated in the reference plane 
$(\Lambda_{Re}^{(\alpha)},\Lambda_{Im}^{(\alpha)} )$. Once the  reaction parameters $c_1$ and $c_2$ have been assigned, one can calculate the coefficients $C_{1,q} (q=0, 1, 2)$ and $C_{2,q} (q=0, ..., 4)$ via Eqs.~(\ref{eq:coeff}). The inequality~(\ref{eq:stabilityReteGen}) allows us to draw the domain of instability, depicted as a shaded region in Fig.~\ref{fig_CGLstabReg_after}. Each eigenvalue (blue circles) of the discrete Laplacian corresponds to a localized point in the plane $(\Lambda_{Re}^{(\alpha)},\Lambda_{Im}^{(\alpha)} )$. 
The instability develops when at least one non-null eigenvalue enters the shaded region.  For an undirected graph,  the points are distributed on the (horizontal) axis, thus outside the region deputed to the instability. When the graph turns asymmetric the imaginary component of $\Lambda^{(\alpha)}$ promotes an instability, which bears a direct imprint of the network topology. As usual, the instability will eventually unfold complex patterns, in the nonlinear regime of the evolution. 

Starting from this setting, and to restore the synchronization,  one can rewire the network connections,  according to the control procedure outlined in the preceding Section \footnote{Remember that the mathematical proofs provided in the previous section hold in general, assuming a directed (balanced and diagonalizable) Laplacian}. In this case, one needs to operate in the complex plane $(\Lambda_{Re}^{(\alpha)},\Lambda_{Im}^{(\alpha)} )$, and act simultaneously on the imaginary component of $\Lambda^{(\alpha)}$, to force the eigenvalues outside the region of instability \footnote{The control can be implemented in different ways. The eigenvalues can be moved for instance horizontally, by acting on their real component, vertically by modifying their imaginary part, or diagonally, by resorting to a linear combination of the two aforementioned strategies. Real eigenvalues lay on the horizontal axis: when falling in the region of instability, they are transferred into the stable domain by sliding them horizontally, namely  by imposing a real correction, the imaginary part proving, in this respect, useless.} . In other words the elements of the diagonal matrix $\mathbf D'$ which encodes for the imposed shifts are, in general, complex. For the case at hands, the spectrum of the controlled  Laplacian operator is displayed in Figs.~\ref{fig_disp_rel_directed} and \ref{fig_CGLstabReg_after} with (red online) stars. The dispersion relation $\lambda_{Re}$ is now consistently negative, reflecting the fact that stabilization has been enforced into the model. Similarly, stars populate the domain of stability in Fig.~\ref{fig_CGLstabReg_after} without invading the shaded portion of the plane.

\begin{figure}
    \includegraphics[width=9cm]{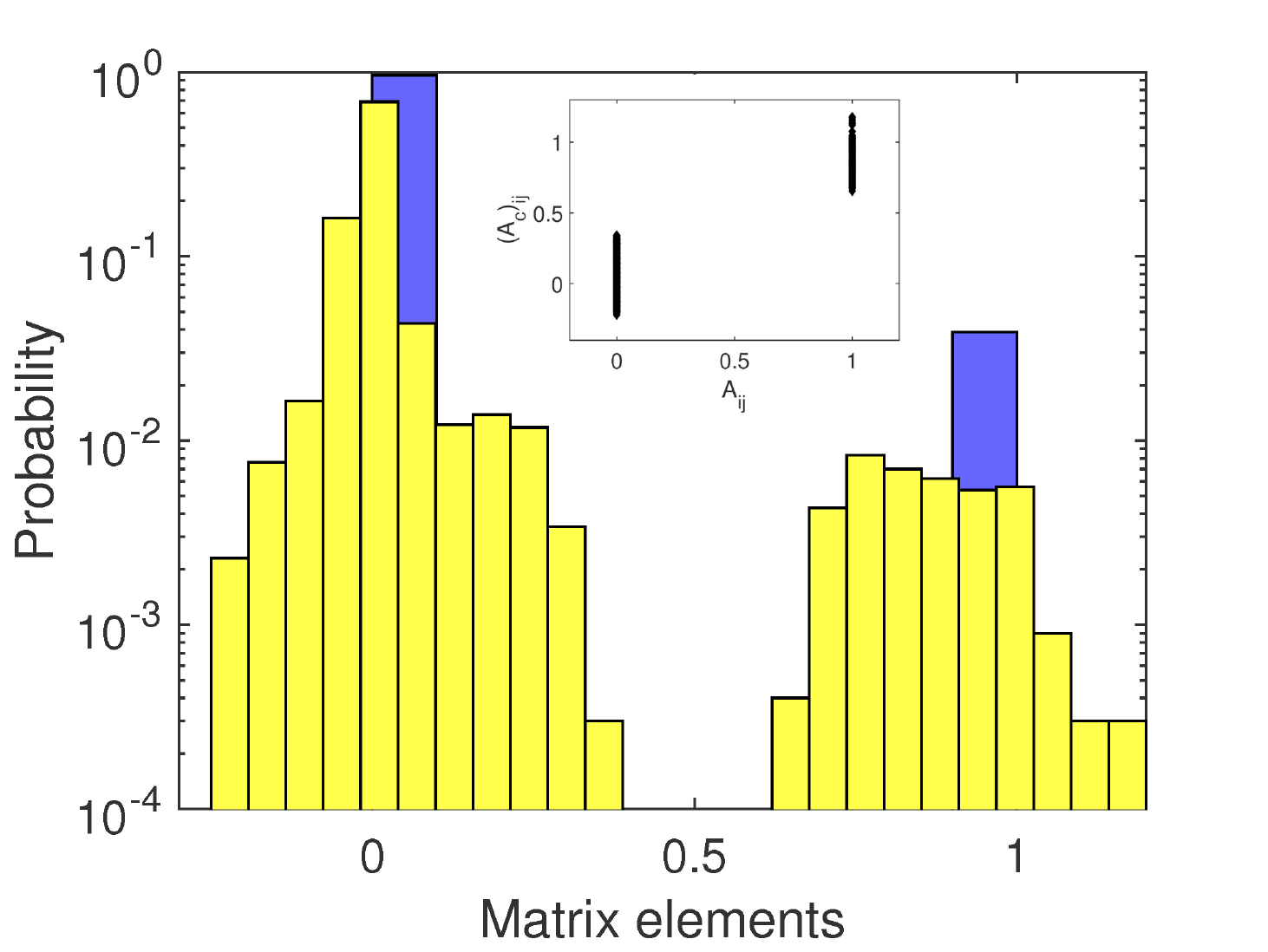}   
   \caption{Main panel: distribution of the elements of the directed NW adjacency matrix before (large dark bins) and after (light small bins) the control. The distribution displays initially two peaks at $0$ and $1$. The controlled adjacency matrix is still bimodal, but the peaks broaden. Importantly, the coupling constant takes also negative values: inhibitory loops should be accommodated for in the rewired weigthed network. Inset: the elements of the controlled adjacency matrix $(A_c)_{ij}$ are plotted vs. the original adjacency matrix  $A_{ij}$. Also for the directed case, a clear correlation between the two is observed. The control induces a rather local modification of the couplings, strong (resp. weak) couplings being preserved under the imposed rewiring.  
}
 \label{isto_A_log_asim}
 \end{figure}

As for the preceding case, the initial adjacency matrix is assumed binary.  The elements of the controlled matrix still display a bimodal distribution (see Fig.~\ref{isto_A_log_asim}): inhibitory coupling are at play as for the case of a symmetric support. 

\begin{figure}[h]
    \vspace{-5cm}
    \hspace*{-3cm}
     \includegraphics[width=13.5cm]{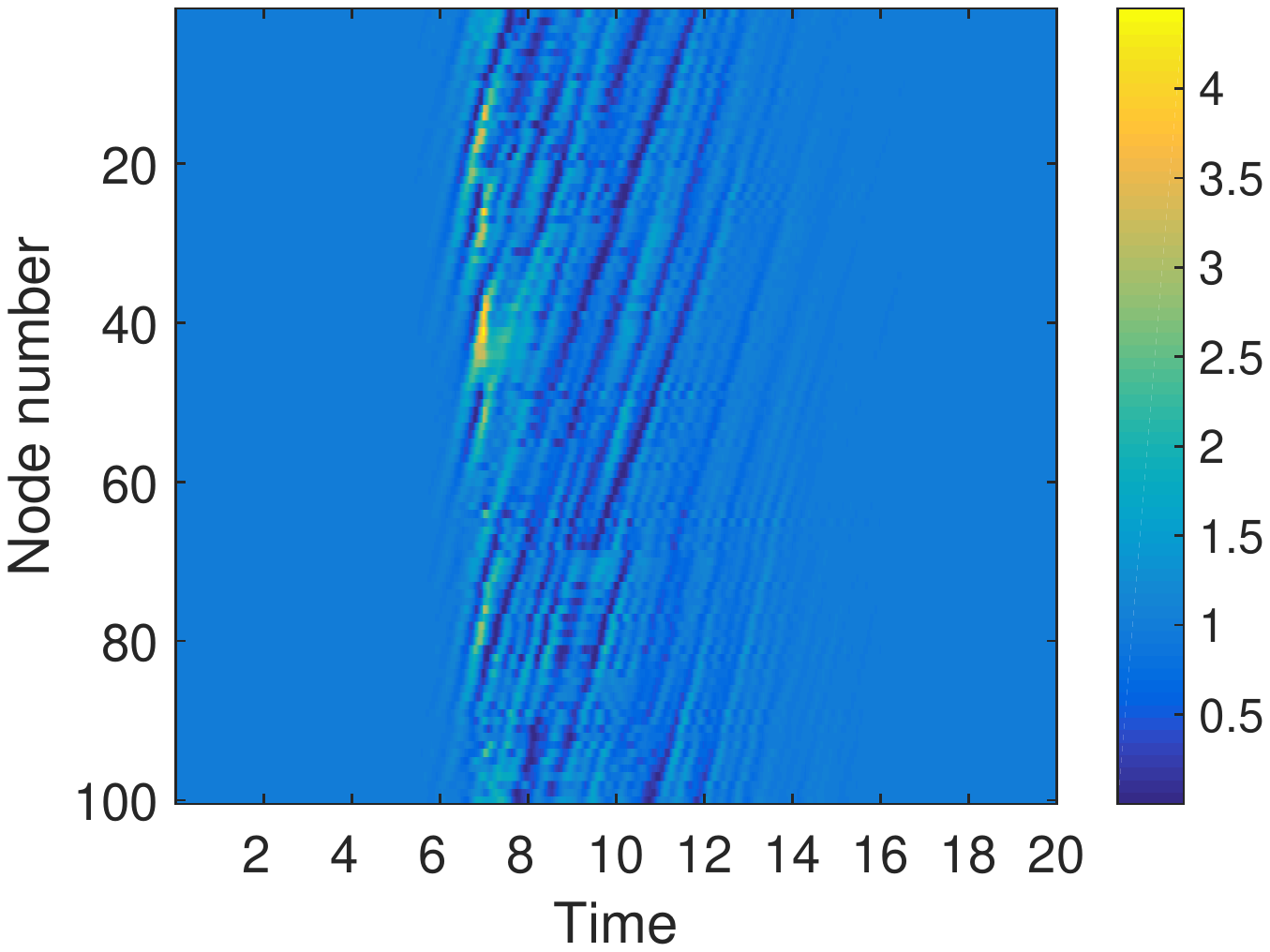}
     \vspace{-6cm}
   \caption{$|\mathbf W|$ vs. time. The system assumes an initially binary (directed and balanced) matrix of connections, as in Fig.~\ref{fig_disp_rel_directed}, and it is unstable to external non homogenous perturbation. At time $\tau_1=5$  the LC is perturbed, and the injected disturbances develop, yielding a loss of synchronization, as predicted by the linear stability analysis. At time $\tau_2=7$ the adjacency matrix is instantaneously controlled, as follows the devised scheme. The perturbation is consequently re-absorbed and the synchronized configuration recovered.}
   \label{fig_CGL_asym_3behav_pert_mod}
  \end{figure}

 To conclude this section  we provide a numerical validation of the implemented method.  In Fig.~\ref{fig_CGL_asym_3behav_pert_mod} we initially evolve the perturbation assuming the unstable and directed adjacency matrix. Then, when the perturbation has evolved in a nonlinear {\it quasi}-wave, the Laplacian is  instantaneously mutated into its controlled counterpart. The perturbation damps and the system regains the initial homogenous consensus state. We again remark that the control is also effective when acted far from the linear regime of the evolution, when nonlinearities are presumably playing a role.\\ 
 
 As a final mandatory remark, we emphasize that the developed control strategy holds in general, beyond the application to the CGLE here considered for purely pedagogical reasons. Indeed, a formally identical scheme can be applied to stabilizing homogenous time-independent fixed points, so preventing the classical Turing-like route to patterns to eventually take place. This extension is discussed for completeness in the Appendix~\ref{app_2rGL}, by employing an {\it ad hoc} multispecies framework which takes inspiration from the Ginzburg-Landau reference model.

  \section{Conclusions}

Patterns are ubiquitous in nature and arise in different contexts, ranging from chemistry to physics, passing through biology and life sciences. The paradigmatic approach to pattern formation deals with a set of reaction-diffusion equations: an initial homogenous equilibrium, constant or time-dependent, can turn unstable via a symmetry breaking instability, instigated by the external injection of a non homogenous disturbance. A non trivial interplay between reaction and diffusion terms, first imagined by Alan Turing in his seminal paper on morphogenesis, is ultimately responsible for the growth of the imposed perturbation. This event takes place for specific choices of the parameter setting and preludes the outbreak of the fully developed patterns.
When the reaction-diffusion system is hosted on a network support, the inherent discreteness and the enforced degree of imposed asymmetry matter in determining the conditions that make the route to patterns possible. The vital role which is played by the topology of the underlying networks of contacts can be efficiently exploited to control the instability and so contrast the drive to pattern formation. In this paper we have elaborated along these lines by devising a suitable control strategy that enforces stabilization, via a supervised  redefinition of the inter-nodes couplings. The idea is to modify the spectrum of the Laplacian by altering the matrix of connections so as to confine the active modes outside the region of instability. The method builds on the work of Nakao~\cite{Nakao14} who numerically showed that an effective stabilization can be achieved by link-rewiring. As in Ref.~\cite{Nakao14}, the Complex Ginzburg-Landau equation has been here assumed as a reference model, to provide a probing test for the newly proposed approach. In this case the control stabilizes the synchronous limit cycle uniform solution. A multispecies system that couples together two real Ginzburg-Landau equations has been also considered, to assess the performance of the method in presence of a homogeneous stationary stable fixed point. When the adjacency matrix is symmetric, the discrete points that constitute the unstable portion of the dispersion relation are moved along the continuum parabola which embodies the characteristic of stability in the idealized continuum limit. Conversely, when the connections are asymmetric, though balanced, different strategies can be implemented to achieve the sought stabilization. One can in general act on the imaginary and real components of the spectrum of the Laplacian operator,  integrating such independent moves as desired. Numerical checks confirmed the effectiveness of the proposed scheme. We recall however that a {\it homogenous}, fixed or time-dependent solution for the system to be controlled, is assumed to exist, which in turn implies dealing with a specific choice for the inter-nodes couplings. The method can be however extended so as to account for the stabilization of  {\it non homogenous equilibria}, as we will report in a separate paper.
   
 \appendix
 \section{Appendix - Stabilizing a homogenous fixed points: multispecies (real) Ginzburg-Landau equations}
 \label{app_2rGL}

This Appendix aims at testing the proposed control method for a multispecies model that undergoes a Turing-like instability. More specifically, we shall consider a 
reaction-diffusion model, hosted on a network, either symmetric or directed (and balanced). Diffusion is governed by the discrete Laplacian operator as introduced in the main body of the paper. The model admits a stationary stable homogeneous fixed point. This latter can turn unstable as follows the injection of a non homogeneous perturbation. In analogy with the discussion carried out for the CGLE, we will consider in a first place the usual setting, the instability resulting from the reactive component of the dynamics (on a symmetric support). Then we will turn to examine the case of a topology-driven instability (directed network). In ideal continuity with the above, we shall assume a specific reaction-diffusion system, consisting of a pair of coupled real Ginzburg-Landau equations, one for each interacting species, $\mathbf{x}$ and $\mathbf{y}$. 
In formulae:

 \begin{equation}
 \left\{\begin {array}{ll} 
  \dot x_i=f(x_i,y_i)+\sum_j\Delta_{ij}x_j\\
  \dot y_i=g(x_i,y_i)+d\sum_j\Delta_{ij}y_j\\
\end{array}\right.
\label{GLreal}
 \end{equation}
 with
 \begin{equation}
  f(x_i,y_i)=\gamma[b_1x_i-(x_i^2+a_1y_i^2)x_i]
 \end{equation}
 and
 \begin{equation}
  g(x_i,y_i)=\gamma[b_2y_i-(x_i^2+a_2y_i^2)y_i],
 \end{equation}
where $x_i$ and $y_i$ are real and positive and $i=1,...N$, with $N$ denoting the number of nodes. The parameters $a_1$, $a_2$, $b_1$, $b_2$, $\gamma$, $d$ are also assumed to be real. As usual $\mathbf{\Delta}$ stands for the discrete Laplacian operator.
System~(\ref{GLreal}) admits an homogenous fixed point  $(\mathbf{x}^*,\mathbf{y}^*)$ whose components respectively read:

 \begin{equation}
 x_i^*=\frac{a_1b_2-a_2b_1}{a_1-a_2}
 \end{equation}
 \begin{equation}
 y_i^*=\frac{b_1-b_2}{a_1-a_2}.
 \end{equation}

 \begin{figure}[t]
   \vspace{-7cm}
   \hspace*{-3cm}
     {\includegraphics[width=14cm]{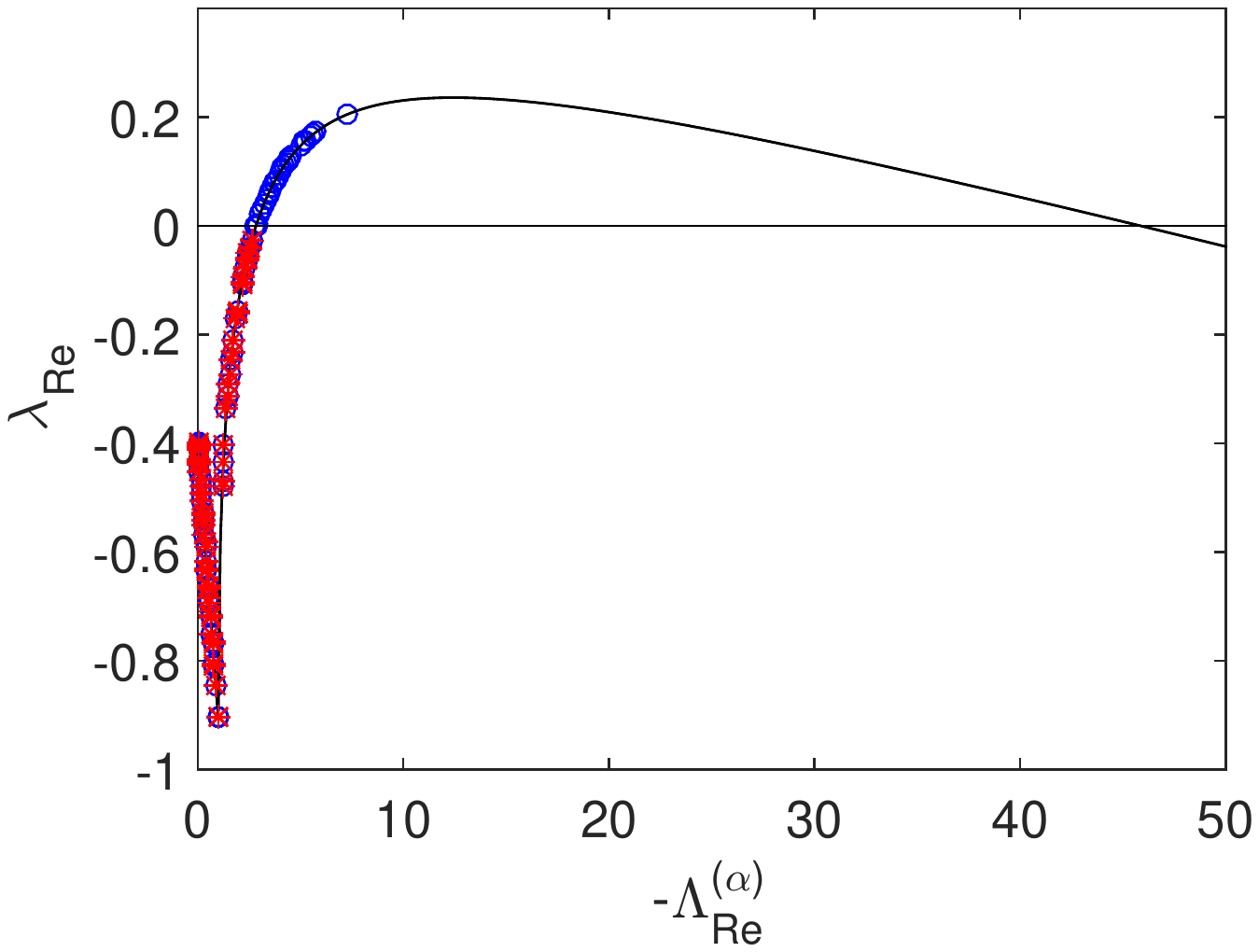}}
   \vspace{-7cm}
   \caption{ The dispersion relation $\lambda_{Re}$ as a function of $-\Lambda_{Re}^{(\alpha)}$. The solid line stands for the continuous case, while 
   symbols refer to the discrete (symmetric) case. Circles (blue online) represent the initial setting, stars (red online) are instead obtained after the control has been applied. Here, $a_1=-3$, $a_2=-1$, $b_1=-1$, $b_2=4$, $\gamma=0.1$ and $d=0.01$. The network employed is identical to that used in drawing Fig.~\ref{fig_CGL_sym_relDisp_before}.}
   \label{fig_RGL_sym_relDisp_before}
  \end{figure}

For a proper choice of the involved parameters, the homogenous fixed point is stable (the detailed study of the stability of the homogenous solution is here omitted). Starting from this setting we insert a non homogenous perturbation in the form $x_i=x^*+u_i$, $y_i=y^*+v_i$ and  linearize Eq.~(\ref{GLreal})  around  $(\mathbf{x}^*,\mathbf{y}^*)$. 
The obtained linear system can be solved by expanding the perturbation on eigenvectors basis, once the underlying network has been specified. This enables one to isolate the region of parameters for which the diffusion driven instability can develop.

  \begin{figure}[t]
  \vspace*{-6.5cm}
  \hspace*{-2.3cm}
   \includegraphics[width=13.5cm]{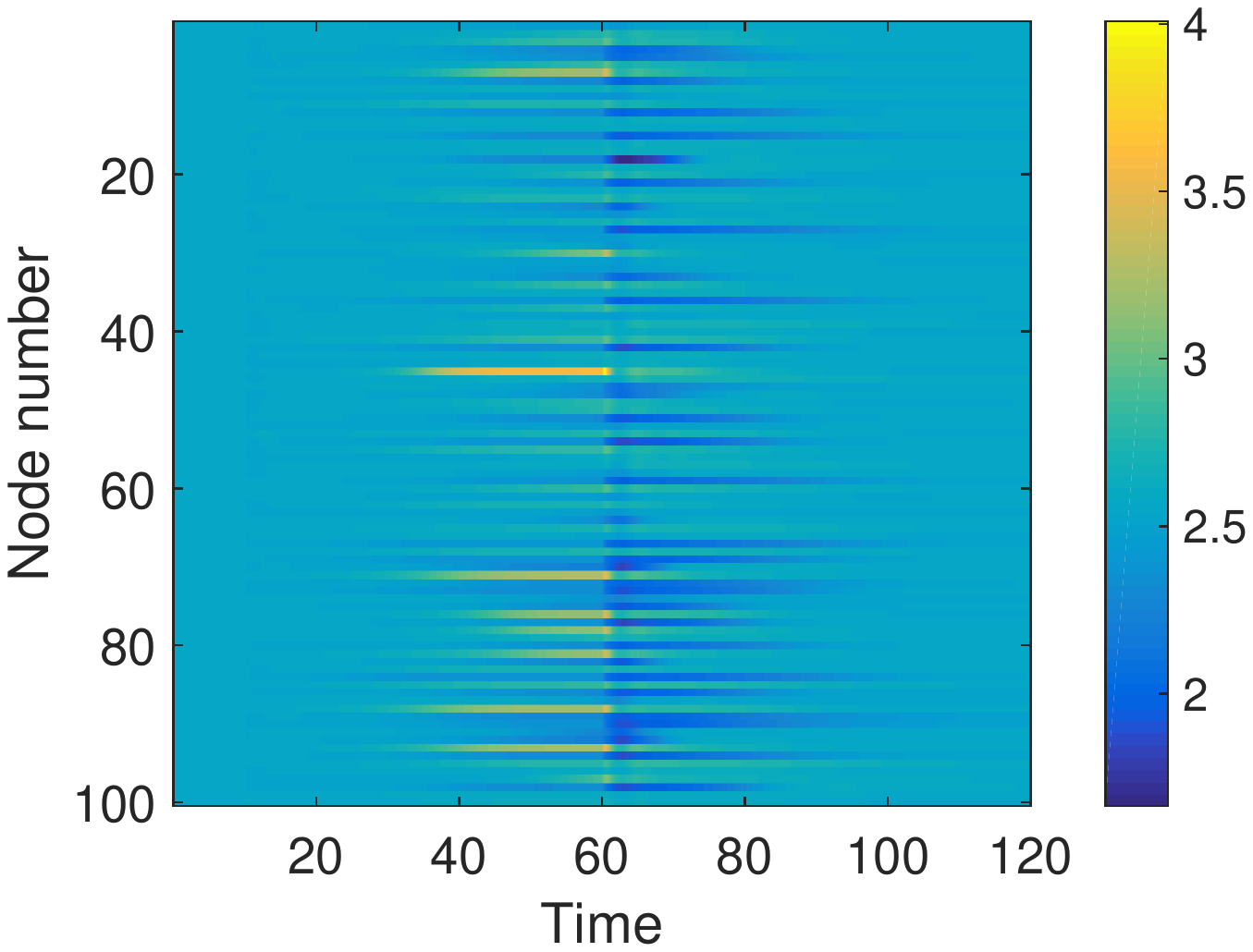}\\
   \vspace*{-12cm}
   \hspace*{-2.3cm}
   \includegraphics[width=13.5cm]{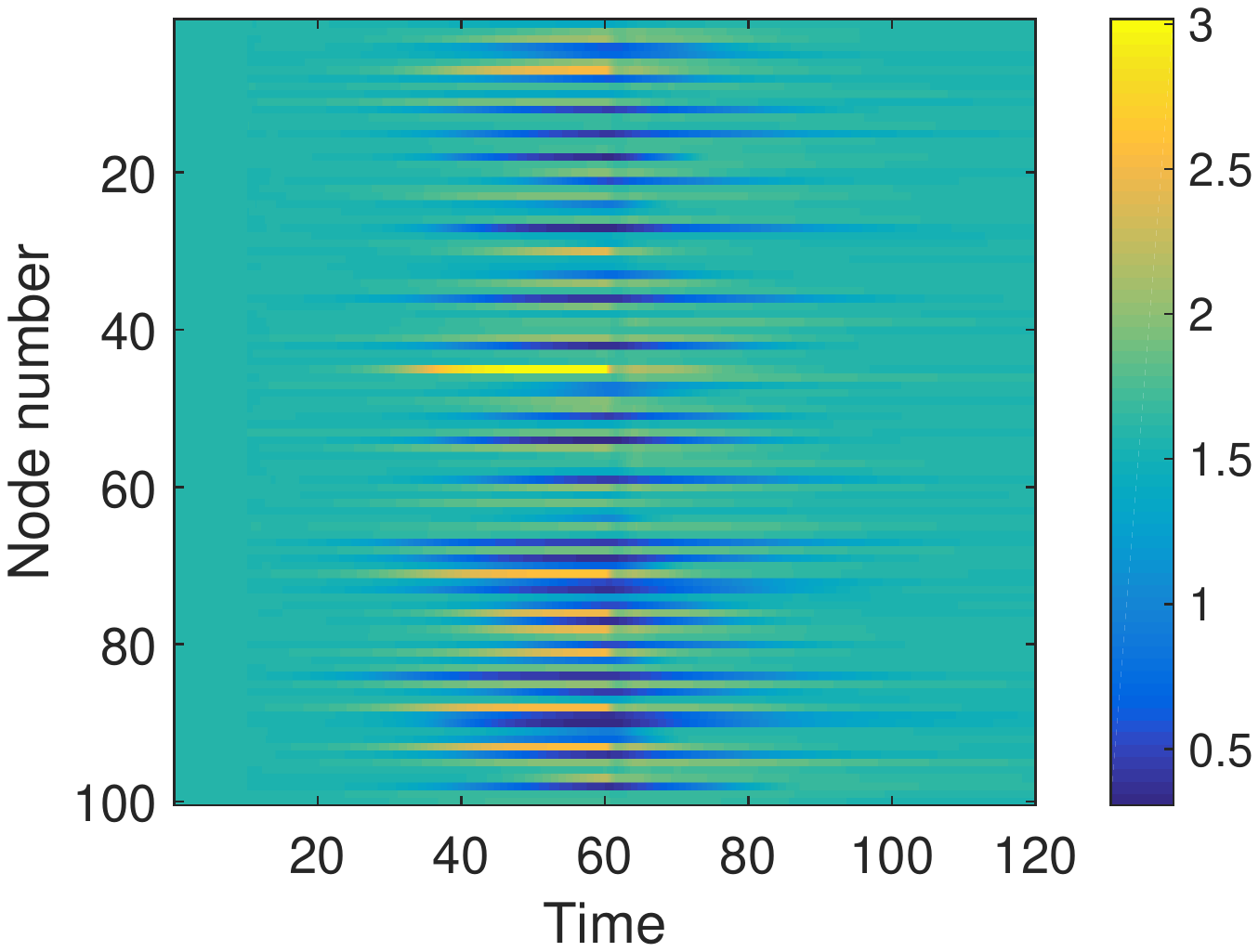}
   \vspace{-6.5cm}
   \caption{Evolution of $\mathbf{x}$ (upper panel) and $\mathbf{y}$ (lower panel) versus time. The system is unstable and a non homogenous perturbation, inserted at time $\tau_1=10$, evolves in patchy distribution. At $\tau_2=60$ the control is turned on (thus the network of connections rewired) and the perturbation damps away. Parameters are set as in Fig.~\ref{fig_RGL_sym_relDisp_before}.}. 
  \label{patt1}
 \end{figure}

In Fig.~\ref{fig_RGL_sym_relDisp_before}, we report the dispersion relation $\lambda_{Re}$ vs. $-\Lambda_{Re}^{(\alpha)}$ for a proper selection of the model parameters.  The solid line stands for the continuum case, while the circles (blue online) are obtained assuming  a (symmetric) network of couplings. Notice that $\lambda_{Re}(\Lambda_{Re}^{(\alpha)}=0)<0$, as expected because of the imposed stability of the homogeneous solution with respect to homogeneous perturbation. Moreover, the circles populate the positive bump of the dispersion relation, thus signaling the instability. The stars (red online) show the dispersion relation computed from the controlled Laplacian. The symbols are now characterized by  $\lambda_{Re}<0$: the stability is hence recovered. Direct simulations of system~(\ref{GLreal}) as displayed in Fig.~\ref{patt1}
 confirm the adequacy of the proposed  control scheme.

\begin{figure}[t]
   \vspace{-7cm}
   \hspace*{-3cm}
     {\includegraphics[width=14cm]{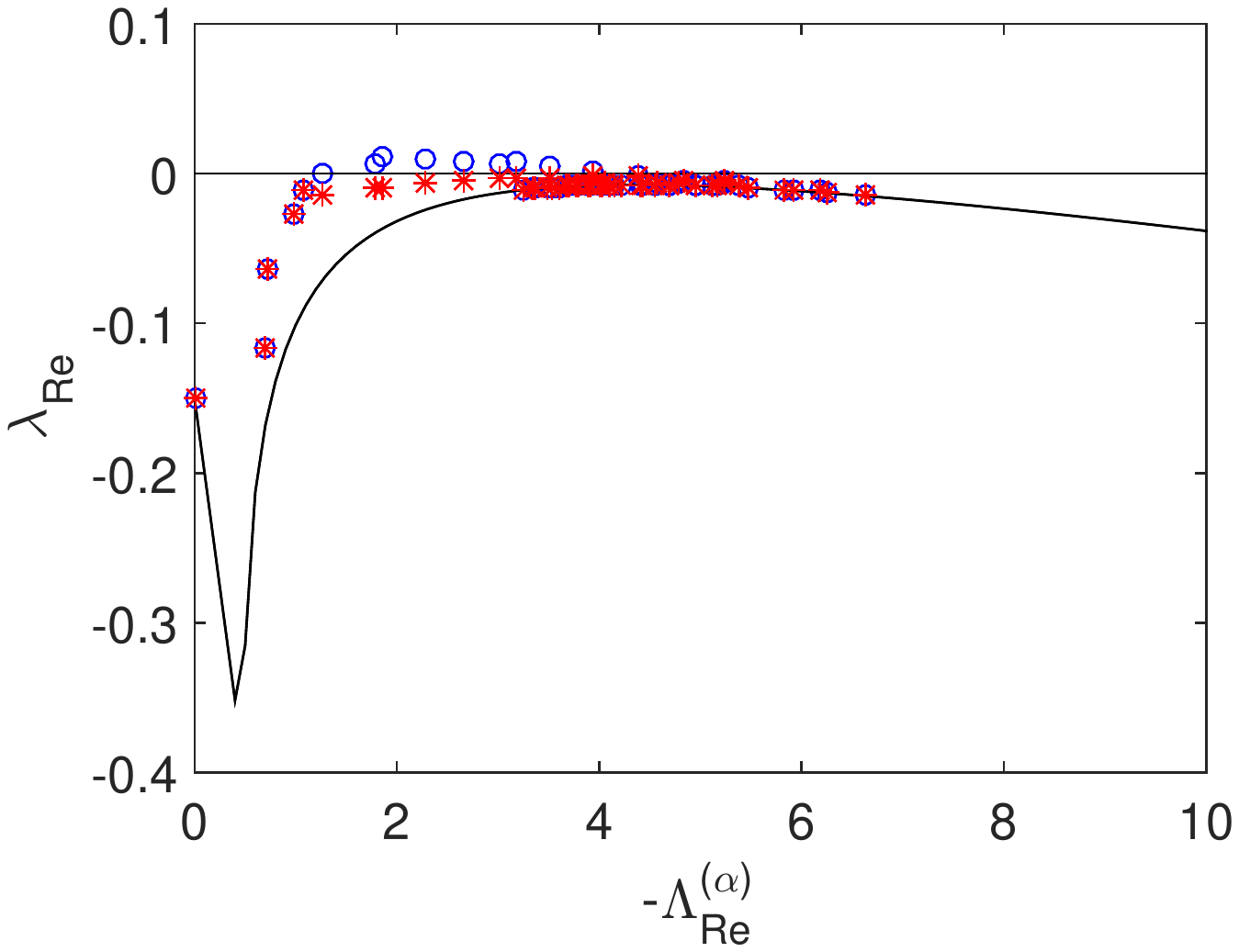}}
   \vspace{-6.8cm}
   \caption{The dispersion relation $\lambda_{Re}$ as a function of $-\Lambda_{Re}^{(\alpha)}$. The solid line stands for the continuous dispersion relation. The (blue online) circle are obtained for the system defined on a NW balanced newtork, with $p=0.27$. The (red online) stars represent the dispersion relation obtained for the controlled matrix of couplings. Here, $a_1=-7$, $a_2=-1$, $b_1=-1$, $b_2=1.5$, $\gamma=0.1$ and $d=0.01$.}
      \label{fig_RGL_asym_relDisp_after}
  \end{figure}
 
  \begin{figure}[t]
   \vspace{-6.75cm}
   \hspace*{-3cm}
     {\includegraphics[width=14cm]{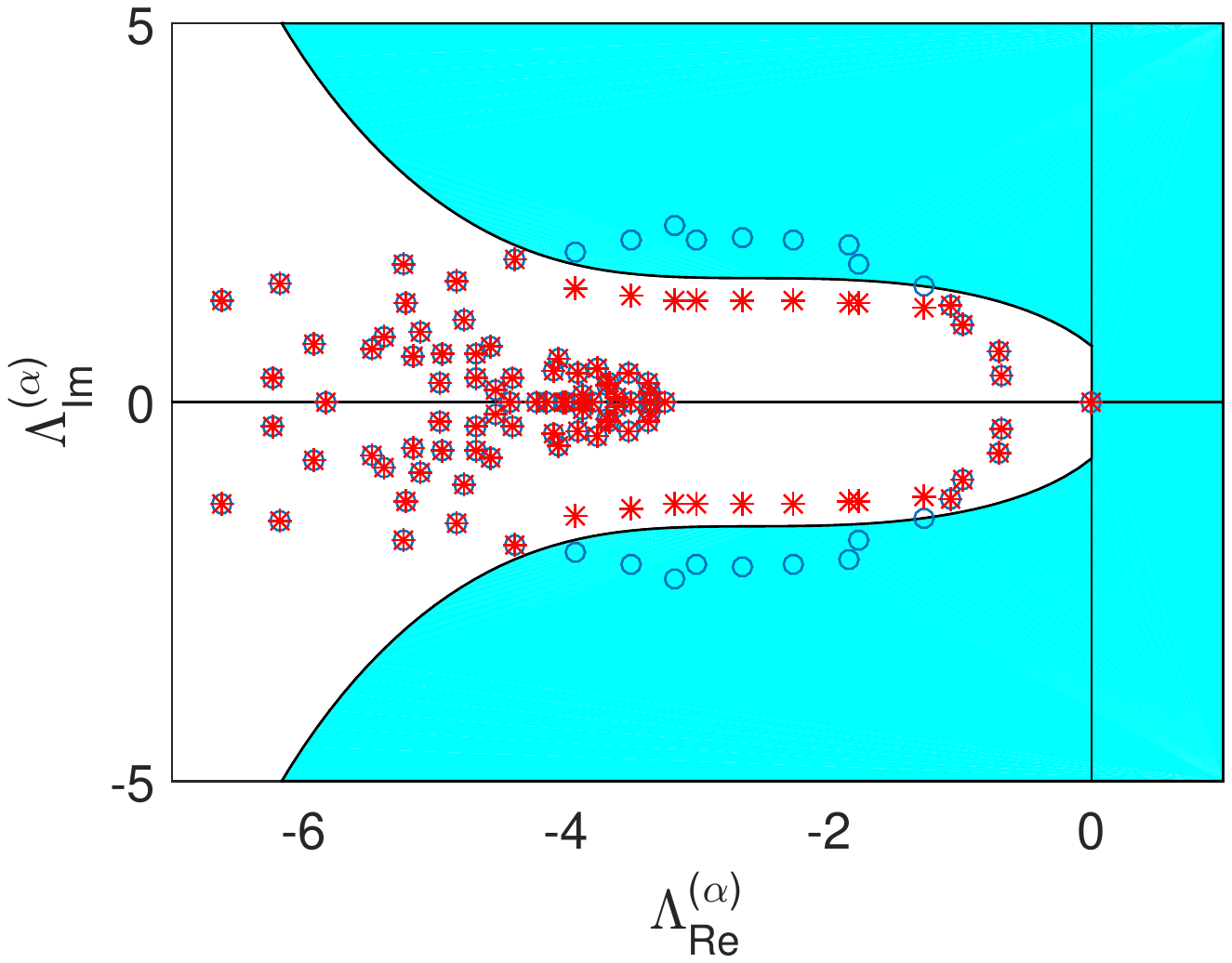}}
   \vspace{-6.7cm}
   \caption{Eigenvalues in the complex plane $(\Lambda_{Re}^{(\alpha)},\Lambda_{Im}^{(\alpha)})$. The blue circles represent the eigenvalues of the initial Laplacian $\mathbf\Delta$, while the red stars refer to the eigenvalues of $\mathbf\Delta_{c}$. The shaded area represents the instability region obtained by using the procedure discussed in Ref.~\cite{AsllaniChallengerPavoneSacconiFanelli14}. The eigenvalues in this region correspond to the unstable modes, characterized by  $\lambda_{Re}>0$,  in Fig.~\ref{fig_RGL_asym_relDisp_after}. Parameters are set as in Fig.~\ref{fig_RGL_asym_relDisp_after}.}   \label{fig_RGLstabReg_after}
  \end{figure}

\begin{figure}[t]
  \vspace*{-6.5cm}
  \hspace*{-2.3cm}
   \includegraphics[width=13.5cm]{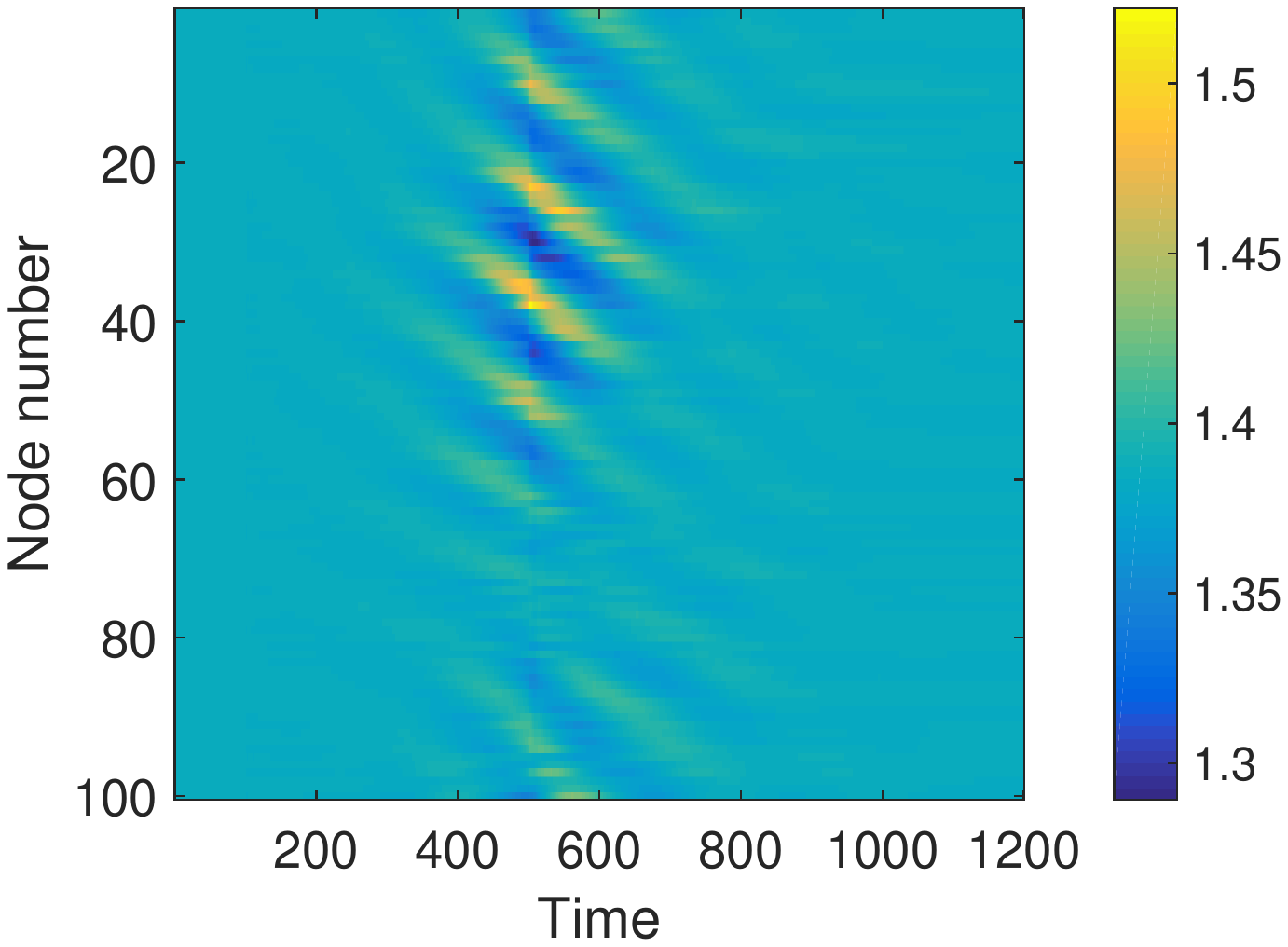}\\
   \vspace*{-12cm}
   \hspace*{-2.3cm}
   \includegraphics[width=13.5cm]{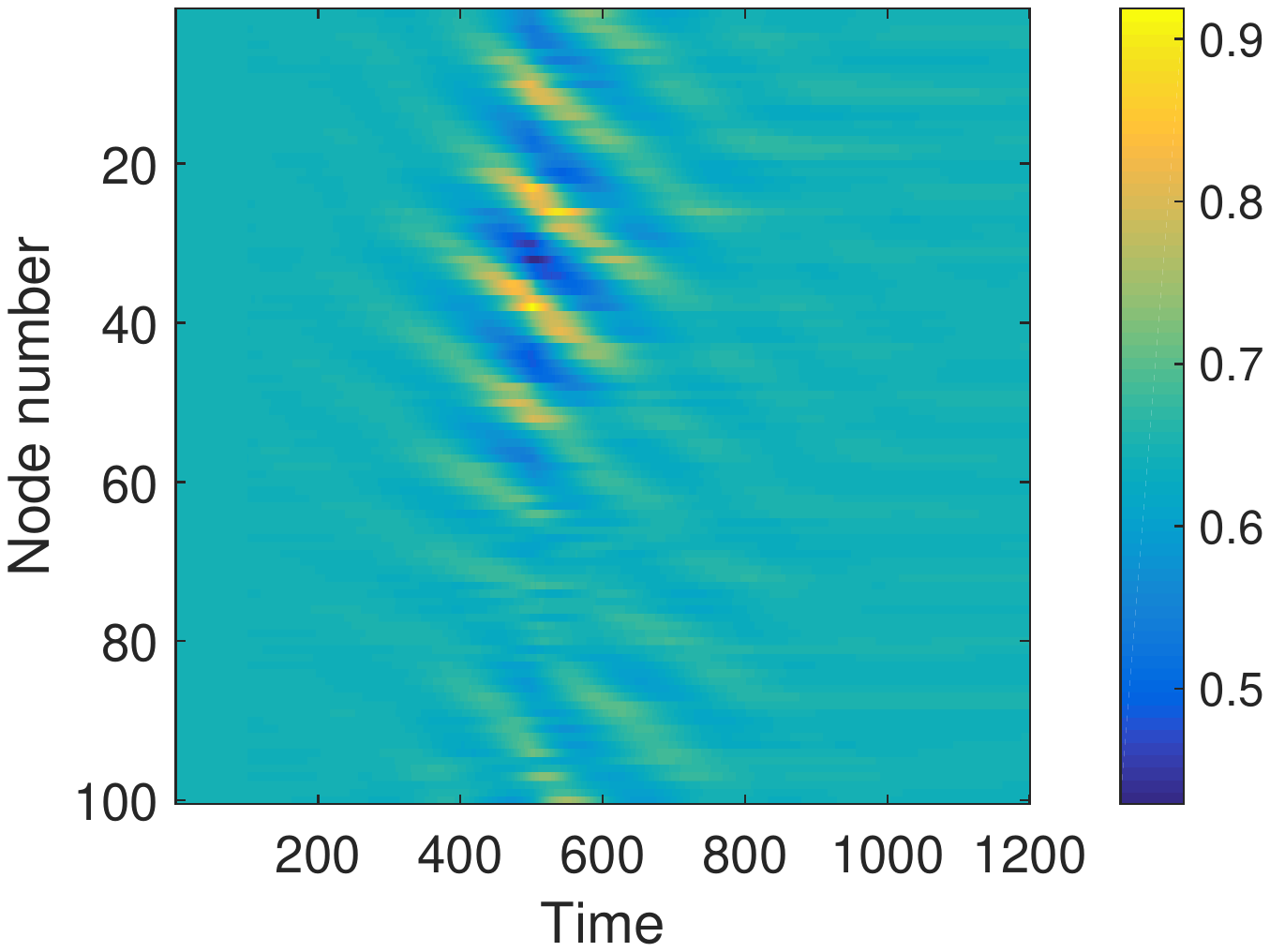}
   \vspace{-6.5cm}
   \caption{Evolution of $\mathbf{x}$ (upper panel) and $\mathbf{y}$ (lower panel) versus time, for the situations depicted in Figs.~\ref{fig_RGL_asym_relDisp_after}
   and \ref{fig_RGLstabReg_after}. The asymmetry in the couplings makes the system unstable and the non homogeneous perturbation inserted at time $\tau_1=100$ evolves in a quasi wave distribution. At $\tau_2=500$ the control is turned on (thus the network of connections rewired) and the perturbation fades away.}
 \label{fig_RGL_asym_evol}
 \end{figure}

We turn then to considering Eq.~(\ref{GLreal})  placed on a directed and balanced network, generated according to the generalized NW recipe discussed in the paper. In Fig.~\ref{fig_RGL_asym_relDisp_after} we report the dispersion relation before (blue circles) and after (red stars) the supervised rewiring of the adjacency matrix. As it can be clearly appreciated by visual inspection, the spectrum of the modified Laplacian matrix yields a stable dispersion relation. The same conclusion can be drawn by analyzing the data reported in Fig.~\ref{fig_RGLstabReg_after}: the (red online) stars are scattered outside the region deputed to the instability, obtained by using the conditions reported in Ref.~\cite{AsllaniChallengerPavoneSacconiFanelli14}. Numerical simulations displayed in Fig.~\ref{fig_RGL_asym_evol}
 confirm the predicted scenario.

\addcontentsline{toc}{chapter}{Bibliography}
\bibliography{myBib}{}
\bibliographystyle{unsrt} 

\end{document}